\documentclass[12pt]{article}

\usepackage[utf8]{inputenc}
\usepackage{textcomp}
\usepackage{amsmath}
\usepackage[shortlabels]{enumitem}
\usepackage{chetnew}
\usepackage{tikz}
\usepackage{amssymb,amsfonts,mathrsfs,dsfont,yfonts,bbm}\usetikzlibrary{calc}
\usepackage{xcolor}
\usepackage{braket}
\usepackage[normalem]{ulem}
\usepackage{import}
\usepackage{booktabs}
\usepackage{varioref}
\usepackage{verbatim}

\newcommand{\cI}{\mathcal{I}}

\newcommand{\sO}{\mathscr O}

\newcommand{\beqn}{\begin{eqnarray}}
\newcommand{\eeqn}{\end{eqnarray}}

\newcommand{\be}{\begin{equation}}
\newcommand{\ee}{\end{equation}}
\newcommand{\bea}{\begin{eqnarray}}
\newcommand{\eea}{\end{eqnarray}}

\newcommand{\CA}{\mathcal{A}}
\newcommand{\CH}{\mathcal{H}}

\newcommand{\CD}{\mathcal{D}}
\newcommand{\CE}{\mathcal{E}}

\newcommand{\CB}{\mathcal{B}}
\newcommand{\CC}{\mathcal{C}}

\newcommand{\CO}{\mathcal{O}}

\newcommand{\CI}{\mathcal{I}}
\newcommand{\CN}{\mathcal{N}}

\newcommand*{\boxcoloro}{orange}
\makeatletter
\newcommand{\boxedo}[1]{\textcolor{\boxcoloro}{%
\tikz[baseline={([yshift=-1ex]current bounding box.center)}] \node [rectangle, minimum width=1ex,rounded corners,draw] {\normalcolor\m@th$\displaystyle#1$};}}
 \makeatother
\newcommand*{\boxcolorr}{red}
\makeatletter
\newcommand{\boxedr}[1]{\textcolor{\boxcolorr}{%
\tikz[baseline={([yshift=-1ex]current bounding box.center)}] \node [rectangle, minimum width=1ex,rounded corners,draw] {\normalcolor\m@th$\displaystyle#1$};}}
 \makeatother
\newcommand*{\boxcolorb}{blue}
\makeatletter
\newcommand{\boxedb}[1]{\textcolor{\boxcolorb}{%
\tikz[baseline={([yshift=-1ex]current bounding box.center)}] \node [rectangle, minimum width=1ex,rounded corners,draw] {\normalcolor\m@th$\displaystyle#1$};}}
 \makeatother
\newcommand*{\boxcolorg}{green}
\makeatletter
\newcommand{\boxedg}[1]{\textcolor{\boxcolorg}{%
\tikz[baseline={([yshift=-1ex]current bounding box.center)}] \node [rectangle, minimum width=1ex,rounded corners,draw] {\normalcolor\m@th$\displaystyle#1$};}}
 \makeatother
 \newcommand*{\boxcolorp}{purple}
\makeatletter
\newcommand{\boxedp}[1]{\textcolor{\boxcolorp}{%
\tikz[baseline={([yshift=-1ex]current bounding box.center)}] \node [rectangle, minimum width=1ex,rounded corners,draw] {\normalcolor\m@th$\displaystyle#1$};}}
 \makeatother
  \newcommand*{\boxcolorc}{cyan}
\makeatletter
\newcommand{\boxedc}[1]{\textcolor{\boxcolorc}{%
\tikz[baseline={([yshift=-1ex]current bounding box.center)}] \node [rectangle, minimum width=1ex,rounded corners,draw] {\normalcolor\m@th$\displaystyle#1$};}}
 \makeatother
  \newcommand*{\boxcolory}{yellow}
\makeatletter
\newcommand{\boxedy}[1]{\textcolor{\boxcolory}{%
\tikz[baseline={([yshift=-1ex]current bounding box.center)}] \node [rectangle, minimum width=1ex,rounded corners,draw] {\normalcolor\m@th$\displaystyle#1$};}}
 \makeatother

\usepackage{amsmath}	

\begin{document}
\preprint{QMUL-PH-22-17}

\title{On the Protected Spectrum of \\[3mm]  the Minimal Argyres-Douglas Theory
 }

\author{Chinmaya Bhargava,$^{\spadesuit}$ Matthew Buican,$^{\diamondsuit}$ and Hongliang Jiang$^{\clubsuit}$}

\affiliation{\smallskip CTP and Department of Physics and Astronomy\\
Queen Mary University of London, London E1 4NS, UK\emails{$^{\spadesuit}$c.bhargava@qmul.ac.uk, $^{\diamondsuit}$m.buican@qmul.ac.uk, $^{\clubsuit}$ h.jiang@qmul.ac.uk}}

\abstract{Despite the power of supersymmetry, finding exact closed-form expressions for the protected operator spectra of interacting superconformal field theories (SCFTs)  is difficult. In this paper, we take a step towards a solution for the \lq\lq simplest" interacting 4D $\CN=2$ SCFT: the minimal Argyres-Douglas (MAD) theory. We present two results that go beyond the well-understood Coulomb branch and Schur sectors. First, we find the exact closed-form spectrum of multiplets containing operators that are chiral with respect to any $\CN=1\subset\CN=2$ superconformal subalgebra. We argue that this \lq\lq full" chiral sector (FCS) is as simple as allowed by unitarity for a theory with a Coulomb branch and that, up to a rescaling of $U(1)_r$ quantum numbers and the vanishing of a finite number of states, the MAD FCS is isospectral to the FCS of the free $\CN=2$ Abelian gauge theory. In the language of superconformal representation theory, this leaves only the spectrum of the poorly understood $\bar\CC_{R,r(j,\bar j)}$ multiplets to be determined. Our second result sheds light on these observables: we find an exact closed-form answer for the number of $\bar\CC_{0,r(j,0)}$ multiplets, for any $r$ and $j$, in the MAD theory. We argue that this sub-sector is also as simple as allowed by unitarity for a theory with a Coulomb branch and that there is a natural map to the corresponding sector of the free $\CN=2$ Abelian gauge theory. These results motivate a conjecture on the full local operator algebra of the MAD theory.\newline}

\medskip

\date{May 2022}

\setcounter{tocdepth}{2}
\maketitle
\toc

\newsec{Introduction}
Inspired by the success of the 2D CFT program initiated by Belavin, Polyakov, and Zamolodchikov (BPZ) almost forty years ago \cite{Belavin:1984vu}, many proposals have been put forward to solve CFTs in $D>2$ (e.g., with various insights building on ideas in \cite{Rattazzi:2008pe} and related works). A slightly more modest, but nonetheless daunting, objective is to find the exact spectrum of operators in such a higher-dimensional CFT. Even more modestly still, finding exact closed-form spectra of \lq\lq protected" operators of interacting superconformal field theories (SCFTs) in $D>2$ is also an important open problem.

In this paper, we make progress toward this last goal in the case of the \lq\lq simplest" interacting 4D $\CN=2$ SCFT: the minimal Argyres-Douglas (MAD) theory discovered a decade after BPZ's groundbreaking work \cite{Argyres:1995jj,Argyres:1995xn}.\footnote{Although we will review in what sense this theory is \lq\lq simplest," we note that it does not have an $\CN=2$ Lagrangian. See \cite{Gadde:2009dj,Gadde:2010zi} for algorithms that in principle allow one to obtain short multiplet spectra in the large-$N$ limit of certain $\CN=2$ Lagrangian theories that are closely related to $\CN=4$ super Yang-Mills. We thank E.~Pomoni for bringing these references to our attention.} In many ways, the MAD theory is the simplest 4D analog of the minimal models studied by BPZ:
\begin{enumerate}
\item The original constructions in \cite{Argyres:1995jj,Argyres:1995xn} find the MAD theory at special points on the Coulomb branches of low-rank $\CN=2$ gauge theories. Therefore, in some sense, the MAD theory has the smallest number of massless \lq\lq ingredients" for an interacting SCFT: an abelian gauge multiplet and two hypermultiplets with mutually non-local electric/magnetic charges.\footnote{Indeed, the maximal simplicity of the MAD theory in the space of 4D $\CN=2$ SCFTs was insightfully suggested by the authors in the conclusions of \cite{Argyres:1995jj}.}
\item The MAD theory has the smallest $c$ central charge of any unitary interacting 4D $\CN=2$ SCFT \cite{Liendo:2015ofa}.\footnote{Although it is not clear if $c$ is a good measure of number of degrees of freedom in 4D $\CN=2$ SCFTs  (it is certainly not for general 4D CFTs, where  the number of degrees of freedom  is measured by the $a$ central charge).}
\item Various sectors of the MAD theory have the simplest possible operator algebra consistent with unitarity and the existence of a Coulomb branch: {\bf(a)} the set of $\CN=2$ chiral operators form a freely generated ring with a single generator \cite{Argyres:1995jj,Argyres:1995xn}. The MAD theory therefore has a one-complex-dimensional Coulomb branch (i.e., the MAD SCFT is \lq\lq rank one"), which is the minimal allowed non-trivial dimension. Moreover, the MAD $\CN=2$ chiral generator has the lowest scaling dimension, $\Delta=\frac65$, among all such generators of rank-1 SCFTs  (this operator dimension implies the theory has no $\CN=2$ Lagrangian; see also the discussion in \cite{Argyres:2015ffa}), {\bf(b)} the set of so-called \lq\lq Schur" operators \cite{Gadde:2011uv} (which includes the energy-momentum tensor and various more complicated operators) is the simplest allowed by 4D unitarity \cite{Buican:2021elx}.

\item Finally, the MAD theory has neither a Higgs branch nor a conformal manifold. In addition, both the 1-form symmetry and the (continuous) 0-form flavor symmetry are trivial \cite{Closset:2020scj,DelZotto:2020esg,Closset:2021lhd,Closset:2021lwy,DelZotto:2022ras,Argyres:2022kon}.\footnote{See \cite{Buican:2021xhs} for a relation between the absence of 1-form symmetry and the absence of an $\CN=2$ conformal manifold.}

\end{enumerate}

The main results of this present work are to expand on point (3) and find maximal simplicity in new sectors of the protected operator spectrum beyond the well-understood $\CN=2$ chiral ring and Schur sector of the MAD theory. This simplicity, along with certain patterns of operator quantum numbers, then lead us to conjecture new structures present in the MAD local operator algebra.

Let us briefly introduce some of our claims in more detail. In order to first set the stage, let us recall the quintessential short multiplet of the 4D $\CN=2$ superconformal algebra (SCA): the $\CN=2$ chiral multiplet. Such a representation has an $\CN=2$ chiral primary, $\sO$, satisfying
\begin{equation}\label{Ochir2}
\left[\bar Q^i_{\dot\alpha},\sO\right]=0~,\ \ \ i=1,2~,
\end{equation}
where $i$ is an $SU(2)_R$ fundamental index, $\dot\alpha$ is a right Lorentz index, and the $\bar Q^i_{\dot\alpha}$ are the four anti-chiral supercharges of the $\CN=2$ SCA. Here $\sO$ is neutral under $SU(2)_R$ but transforms under the $U(1)_r\subset U(1)_r\times SU(2)_R$ superconformal $R$ symmetry (we will denote this latter charge as $r(\sO)\ne0$). In the language of \cite{Dolan:2002zh}, $\sO$ is a primary of an $\bar\CE_{r(0,0)}$ multiplet\footnote{We mostly follow the notation of \cite{Dolan:2002zh} for short multiplets; however, the sign of the $r$ charge is flipped: $r_\text{here} =-r_\text{there}$. On the other hand, as in \cite{Dolan:2002zh}, $R$ is the $SU(2)_R$ weight (and can therefore be negative) except when $R\ge 0$ labels the multiplet. In this case, $R$ is the $SU(2)_R$ spin of the superconformal primary (or, equivalently, the $R$ weight for the $SU(2)_R$ highest-weight component of the superconformal primary).} (see also the equivalent taxonomies in \cite{Dobrev:1985qv,Cordova:2016emh}).

In the case of the MAD theory, the spectrum of these multiplets is well understood: the various $\CN=2$ chiral operators form a ring that is freely generated with generator $\sO$ having $r(\sO)=6/5$. In particular, we have all $\sO^n$ (with $r=6n/5$) and so we have one $\bar\CE_{6n/5(0,0)}$ multiplet for every $n\ge1$. We say that an $\CN=2$ chiral operator is non-trivial in the $\CN=2$ chiral ring if, for any $i$ (or combination thereof)
\begin{equation}
\sO\ne\left\{\bar Q^i_{\dot\alpha},\sO'\right\}~,
\end{equation}
where $\sO'$ is any well-defined local operator. Otherwise, a chiral operator is said to be trivial in the $\CN=2$ chiral ring. Hoping no confusion arises, we will write this condition as $\sO=0$ (even though $\sO$ may be non-vanishing in the local operator Hilbert space). Note that trivial $\CN=2$ chiral ring operators cannot be primaries of $\bar\CE_r$ multiplets.

More generally, it is interesting to study less protected operators than those in \eqref{Ochir2}. One natural class of operators to study are those that are annihilated by half the anti-chiral supercharges. Without loss of generality, we can consider $ \CO$ annihilated by $\bar Q^1_{\dot\alpha}$ 
\begin{equation}\label{N1chidef}
\left[\bar Q^1_{\dot\alpha}, \CO\right\}=0~,
\end{equation}
but not necessarily annihilated by $\bar Q^2_{\dot\alpha}$.
We can again define an $\CN=1$ chiral ring, or a full chiral sector (FCS), by further introducing the equivalence relation $\CO \sim \CO +\left\{\bar Q^1_{\dot\alpha}, \CO'\right]$. Therefore, $ \CO$ is non-trivial in the FCS if and only if 
\be\label{Ntrivial}
  \CO\ne\left\{\bar Q^1_{\dot\alpha}, \CO'\right]~, 
  \ee for  any well-defined local operator $ \CO'$. When  \eqref{Ntrivial} is violated, we will write $ \CO=0$ to indicate triviality in the FCS (even though $ \CO$ can be non-vanishing in the local operator Hilbert space).

Which multiplets can operators such as $ \CO$ of \eqref{N1chidef} and \eqref{Ntrivial} sit in? A quick look at representations of the SCA in \cite{Dolan:2002zh} reveals the following:

\bigskip
\noindent
{\bf Fact:} The only representations of the SCA that can house $ \CO$ satisfying \eqref{N1chidef} and \eqref{Ntrivial} are of the form\footnote{Antichiral operators satisfying the conjugate conditions sit in $\overline{\rm FCS}:=\CE\oplus\CD\oplus\hat\CB\oplus\CB$.}
\begin{equation}\label{FCSdef}
{\rm FCS}:=\bar\CE\oplus\bar\CD\oplus\hat\CB\oplus\bar\CB~.
\end{equation}
We define the FCS to be the set of the above multiplets (along with the corresponding chiral OPEs).   See Appendix~\ref{chiralOPgeneral} for some properties of the FCS in  general 4D $\CN=2$ SCFTs. Note that in the MAD theory it is already known that there are neither $\bar\CD$ nor $\hat\CB$ multiplets (e.g., see \cite{Buican:2021elx}).

\bigskip
\noindent
The list of multiplets in \eqref{FCSdef} can be understood heuristically as follows. These are the only representations that obey shortening conditions that are linear in the various supercharges and involve no contractions of supercharge spin indices. All other short representations (i.e., the $\hat\CC$ and $\bar\CC\oplus\CC$ multiplets) obey shortening conditions that are quadratic in the supercharges or conditions involving contractions of supercharge spins (so-called \lq\lq semi-shortening" conditions in the nomenclature of \cite{Dolan:2002zh}). As a result, such multiplets cannot house a non-trivial element of the FCS because then they would necessarily obey additional shortening conditions.

Given this groundwork, we can state our first main result: in the MAD theory there are no $\bar\CB$ multiplets (and, by CPT, there are no $\CB$ representations either). As a result, the MAD FCS is as simple as allowed by unitarity and the existence of a Coulomb branch
\begin{equation}\label{MADFCS}
{\rm FCS}_\text{MAD}=\bar\CE~.
\end{equation}

Our argument in favor of \eqref{MADFCS} revolves around the RG flow to the MAD theory studied in \cite{Maruyoshi:2016tqk} involving $\CN=1$ in the UV accidentally enhanced to $\CN=2$ in the IR. We start by constructing the naive $\CN=1$ chiral ring generators in the UV. Our key idea is to then demand that these operators form a part of an $\CN=2$ FCS in the IR. Checking this fact amounts to a representation theoretical version of the more analytic arguments on chiral rings in \cite{Cachazo:2002ry}.\footnote{We should also note that we find the same FCS generators as the authors in \cite{Xie:2016hny}, who used an approach inspired by \cite{Cachazo:2002ry} and studied an $\CN=2$ RG flow with broken $U(1)_r$ based on one-flavor $\CN=2$ $SU(2)$ SQCD. We extend this result by proving that there are no $\bar\CB$ multiplets. In addition, our proof involves an RG flow with manifest $U(1)_r\times SU(2)_R$ Cartans. As a result, operator mixing is fully under control in our flow and provides a proof of the claim in \cite{Xie:2016hny}.}

The vanishing of the $\bar\CB$ multiplets then leads to our second result. Indeed, we argue that it is then possible to compute the exact spectrum of $\bar\CC_{0,r(j,0)}$ multiplets for any $r$ and $j$ from the superconformal index (since the index can, a priori, involve highly non-trivial cancellations, this statement is not obvious).

These latter multiplets are of interest for two main reasons. First, the $\bar\CC_{R,r(j,\bar j)}$ multiplets are the last unknown pieces of the MAD short multiplet spectrum. While we restrict ourselves to the subset of multiplets with $R=\bar j=0$, our results shed the first non-perturbative (in index fugacities and quantum numbers) light on this sector. The second reason for being interesting is that they arise in the OPEs of FCS and Schur operators. These latter operators sit in multiplets closely related to those in \eqref{FCSdef}
\begin{equation}
{\rm Schur}:=\CD\oplus\bar\CD\oplus\hat\CB\oplus\hat\CC~.
\end{equation}
The Schur multiplets are well understood since they turn out to be isomorphic (in a sense defined in \cite{Beem:2014zpa}) to operators in a 2D vertex operator algebra (VOA). 

The main point is that our result on the $\bar\CC_{0,r(j,0)}$ spectrum shows that these multiplets act as an OPE-induced bridge between the Schur sector and the FCS. This realization motivates us to conjecture general constraints on the MAD local operator algebra (including for  long multiplets).

In the next section, we build the FCS of the MAD theory. We conclude with a proof of our result on chiral operators. We then proceed to consider $\bar\CC$ operators. Afterwards, we turn our attention to the free Abelian gauge theory and show that both its FCS and $\bar\CC_{0,r(j,0)}$ sectors are closely related to those of the MAD theory. Using this fact, along with our two main results as inspiration, we then proceed to our conjecture.

\newsec{The full chiral ring: beyond the Coulomb branch operators}\label{chirRing}

All interacting 4D $\CN=2$ SCFTs are believed to possess an $\CN=2$-preserving Coulomb branch.\footnote{We define the Coulomb branch to be the set of vacua in which the superconformal $U(1)_r$ is spontaneously broken, and the $SU(2)_R$ is unbroken. Vacua on the Coulomb branch necessarily contain at least one $\CN=2$ free vector multiplet, but they can also contain additional degrees of freedom at generic points (e.g., as in the case of $\CN=4$ SYM).} As discussed in the introduction, this moduli space is parameterized by ($SU(2)_R$-neutral) $\CN=2$ chiral operators (i.e., operators annihilated by all four anti-chiral supercharges) that are, in the nomenclature of \cite{Dolan:2002zh}, $\CN=2$ superconformal primaries of $\bar\CE_r$ multiplets.\footnote{In principle, the superconformal algebra allows $\bar\CE_{r(j,0)}$ multiplets to have $j\ne0$, but a careful analysis of locality shows that $j=0$ in 4D $\CN=2$ SCFTs \cite{Manenti:2019jds} (as conjectured and verified in various classes of theories in \cite{Buican:2014qla}). Therefore, we drop the spin from the label of these multiplets.} The corresponding $\CN=2$ (Coulomb branch) chiral ring is infinite and, in the simplest theories, has a single generator.

In the case of the MAD theory, the single generator of the chiral ring has $\Delta=r=6/5$ (here $\Delta$ is the scaling dimension) and will be denoted as $\CO_{6/5}\in\bar\CE_{6/5}$. The MAD Coulomb branch chiral ring, $\mathcal{R}_{\CC}$, is then described as
\begin{equation}\label{N2chiral}
\mathcal{R}_{\CC}:={\rm Span}\left(\left\{\CO_{6n/5}|n\in\mathbb{Z}_{>0}\right\}\right)~,\ \ \ \CO_{6n/5}=(\CO_{6/5})^n~,
\end{equation}
where we take arbitrary linear combinations of the $\CO_{6n/5}$.

In what follows, we will be interested in studying more general operators that are chiral with respect to an $\CN=1\subset\CN=2$ superconformal subalgebra. In fact, we will see that the MAD theory has, in a sense we will make precise, the simplest set of such operators allowed by unitarity and the existence of a Coulomb branch.

To understand this statement, we first fix an $\CN=1$ subalgebra. Without loss of generality, we consider the subalgebra generated by
\begin{equation}\label{N1sub}
Q_{1\alpha}\sim Q^2_{\alpha}\in\left(  \frac12,0\right)_{-\frac12,-\frac12}~,\ \ \ \bar Q^1_{\dot\alpha}\sim\bar Q_{2\dot\alpha}\in\left(0,  \frac12\right)_{\frac12,\frac12}~,
\end{equation}
where we write the quantum numbers as $(j,\bar j)_{R,r}$, with $j$ the left spin, $\bar j$ the right spin, $R$ the $SU(2)_R$ weight,\footnote{Note that    $SU(2)$ Lorentz spins $j,\bar j\in\frac12 \mathbb N$ are non-negative, while  the  $SU(2)_R$ weight,  $R \in\frac12 \mathbb Z$, can be either positive or negative. Therefore $(j,\bar j)_{R,r}$ actually represents a Lorentz multiplet with specific $R,r$ charge.} and $r$ the $U(1)_r$ charge (note that $R-r$ is a flavor symmetry of this subalgebra).\footnote{To get a superconformal algebra, we should also include $S_{2\alpha}\sim S^1_{\alpha}$ and $\bar S_{1\dot\alpha}\sim\bar S^2_{\dot\alpha}$.}

Then, using the orthogonal algebra generated by 
\be\label{N1perp}
Q_{2\alpha}\sim Q^1_{\alpha}\in\left(  \frac12,0\right)_{ \frac12,-\frac12}~,\ \ \ \bar Q^2_{\dot\alpha}\sim\bar Q_{1\dot\alpha}\in\left( 0,  \frac12\right)_{-\frac12, \frac12}~ ,
\ee  
we see that an $\bar\CE_r$ multiplet contains three chiral primaries with respect to the $\CN=1$ subalgebra in \eqref{N1sub} (note that $F^{\perp}:=R+r$ is a flavor symmetry of the orthogonal subalgebra). In particular, these primaries embed as follows
\be\label{Echiral}
\CO_{r}\in (0,0)_{0,r} \xrightarrow{Q^1_\alpha} \CO_{r,\alpha}\in (  1/2,0)_{1/2, r-1/2} \xrightarrow{ (Q^1)^2:=\epsilon^{\alpha\beta} Q^1_\alpha Q^1_\beta }  \CO'_{r }\in(0,0)_{1,r-1}~,
\ee 
where the leftmost operator is the $\CN=2$ chiral primary, and the remaining operators are level-one and level-two superconformal descendants. Note that $\CO_{r,\alpha}$ and $\CO'_{r}$ are not in the $\CN=2$ chiral ring (they are not annihilated by $\bar Q^2_{\dot\alpha}$), but they are in the $\CN=1$ chiral ring defined by \eqref{N1chidef}. In fact, they are non-trivial elements of the $\CN=1$ chiral ring since the structure of the $\bar\CE_r$ multiplet does not allow them to be $\bar Q^1_{\dot\alpha}$-descendants of another operator in that multiplet. For the MAD theory, we have $r=\frac65n$ with $n\in \mathbb Z_{>0}$.

We now arrive at the main claim of this section:

\bigskip
\noindent
{\bf Claim 1:} The only operators in the MAD theory that are non-trivial chiral ring elements with respect to an $\CN=1\subset\CN=2$ sub-algebra are the three $\CN=1$ chiral operators in each $\bar\CE_{6n/5}$ multiplet ($n\ge1$).

\bigskip
\noindent
A quick scan of the allowed unitary superconformal representations in \cite{Dolan:2002zh} shows that, besides the $\bar\CE_r$ multiplets, the only other multiplets that can host $\CN=1$ chiral operators are of type $\bar\CB_{R,r(j,0)}$, $\bar\CD_{R(j,0)}$, and $\hat\CB_R$; however, recall from \cite{Buican:2021elx} that the MAD theory has neither $\hat\CB_R$ nor $\bar\CD_{R(j,0)}$ multiplets.

The putative $\CN=1$ chiral operators in $\bar\CB_{R,r(j,0)}$ multiplets can be described as $SU(2)_R$ highest-weight primaries and descendants at levels one and two
\bea\label{Bbar}
\CO^{11\cdots1}_{\alpha_1\cdots\alpha_{2j}}\in (j,0)_{R,r} &\xrightarrow{Q^1_\alpha}& \CO_{\alpha_1\cdots\alpha_{2j}\alpha}^{11\cdots11}\oplus\CO_{\alpha_1\cdots\alpha_{2j-1}\alpha}^{11\cdots11}\in (j+\frac12,0)_{R+\frac12,r-\frac12}\oplus(j-\frac12,0)_{R+\frac12,r-\frac12} \cr&\xrightarrow{(Q^1)^2 }&  \CO_{\alpha_1\cdots\alpha_{2j}}^{11\cdots111}\in(j,0)_{R+1,r-1}~,
\eea 
where superscripts denote $SU(2)_R$ fundamental weights, and the $\alpha_i$ are fundamental left spin indices.\footnote{Note that acting with $\bar Q^2_{\dot\alpha}$ gives additional $\CN=1$ chiral operators. However, these are trivial in the $\CN=1$ chiral ring. Indeed, consider $\left[\bar Q^2_{\dot\alpha},\CO^{11\cdots1}_{\alpha_1\cdots\alpha_{2j}}\right\}$. Now recall that the shortening condition of the primary is
\begin{equation}
\left[\bar Q^1_{\dot\alpha},\CO^{11\cdots1}_{\alpha_1\cdots\alpha_{2j}}\right\}=0~.
\end{equation}
Acting with an $SU(2)_R$-lowering operator, we see that $\left[\bar Q^2_{\dot\alpha},\CO^{11\cdots1}_{\alpha_1\cdots\alpha_{2j}}\right\}=-2R\left[\bar Q^1_{\dot\alpha},\CO^{21\cdots1}_{\alpha_1\cdots\alpha_{2j}}\right\}$ is trivial in the $\CN=1$ chiral ring.
} As a result, we have the following corollary:

\bigskip
\noindent
{\bf Corollary 1:} The MAD theory has no $\bar\CB_{R,r(j,0)}$ multiplets (by CPT the same is true for $\CB_{R,-r(0,j)}$).

\bigskip
\noindent
Defining, as in \eqref{FCSdef} of the introduction, the set of $\bar\CE\oplus\hat\CB\oplus\bar\CB\oplus\bar\CD$ multiplets to be the \lq\lq full chiral sector" of a 4D $\CN=2$ SCFT, we see that

\bigskip
\noindent
{\bf Corollary 2:} The full chiral sector (FCS) of the MAD theory, written in \eqref{MADFCS}, is as simple as possible for a unitary theory with a Coulomb branch.\footnote{To understand the importance of unitarity, consider the free $\CN=2$ Abelian gauge theory with fields of wrong statistics (e.g., see the discussions in \cite{Dijkgraaf:2016lym} for more general theories along these lines). In this case, the chiral ring generator, $\varphi$, is nilpotent: $\varphi^2=0$. Therefore, the full chiral sector of this non-unitary theory consists of operators in a single $\bar\CD_{0(0,0)}$ multiplet. Of course, one may object to this theory having a Coulomb branch since the moduli space is just a point. On the other hand, its dynamics is that of an abelian gauge theory (albeit with wrong statistics).} Moreover, if all unitary interacting theories have a Coulomb branch, then the FCS of the MAD theory is as simple as possible for a unitary interacting theory.\footnote{This is the chiral ring equivalent of the minimality of the MAD Schur sector proven in \cite{Buican:2021elx}.}

\subsec{A proof of Claim 1}
  
Let us now give a proof of Claim 1 and therefore also of Corollaries 1 and 2. To that end, our strategy will be to use the $\CN=1$ Lagrangian of \cite{Maruyoshi:2016tqk} (as modified in \cite{Benvenuti:2018bav}; see also \cite{Benvenuti:2017lle} for related discussions). In particular, we consider the $\CN=1$ $SU(2)$ gauge theory with matter content in table \ref{charge} and superpotential
\begin{equation}\label{WUV}
W= X\phi^2+ M\phi q'q'+ \phi q q~,
\end{equation}
where we have suppressed couplings in front of the terms in $W$. We have also defined $\phi^2:=\phi^a\phi^a$, $\phi q'q':=\phi^a(q' q')^a$, and $\phi qq:=\phi^a(qq)^a$ with the $SU(2)$ adjoint index, $a=1,2,3$, summed over. Note that the UV theory is written in terms of the $\CN=1$ supercharges in \eqref{N1sub} (the remaining $\CN=2$ supercharges are emergent in the IR).

\begin{table} [h]
 \begin{center}\renewcommand{\arraystretch}{1.2}
\begin{tabular}{ c|c l   r c } 
 fields & SU(2) &   $r^\text{IR}$ &$I_3^\text{IR}$ \\ \hline
 $q$ & $\square$ & $ \frac25$ &$\frac12$ \\
 $q'$ & $\square$ &  $ -\frac15$  &$\frac12$\\
 $\phi$ & adj. & $ \frac15$ &$0$ \\
 $M$ &   \bf 1 & $ \frac65$  &$0$\\
 $X$ & \bf 1 & $ \frac35$ &$1$\\
  $\lambda_\alpha $ & adj. & $ \frac12$  &$\frac12$\\
 \end{tabular}
 \caption{The charges of primaries of all $\mathcal  N=1$ (chiral) UV fields. The first column of charges give the representation under the $SU(2)$ gauge group. Here  $I_3^\text{IR}$ refers to the $\CN=2$ superconformal $SU(2)_R$ weight in the IR (i.e., $I_3^\text{IR}\to R$), and $r^\text{IR}$ refers to the $\CN=2$ superconformal $U(1)_r$ charge in the IR (i.e., $r^\text{IR}\to r$), which is determined through $a$-maximization. These quantum numbers are visible in the UV description (hence the power of this RG flow), and the superpotential has quantum numbers $r^\text{IR}(W)=I_3^\text{IR}(W)=1$.}
 \label{charge}
\end{center}
 \end{table}

Our algorithm for proving Claim 1 can be summarized as follows:
\begin{enumerate}
\item Write down all possible naive $\CN=1$ chiral ring generators in the UV theory.
\item Demand that, in the IR, all $\CN=1$ chiral ring generators sit in either $\bar\CE_r$ multiplets (as in \eqref{Echiral})  or in $\bar\CB_{R,r(j,0)}$ multiplets (as in \eqref{Bbar}).
\item{Argue that any $\CN=1$ chiral ring operators (including composites built from generators) cannot sit in $\bar\CB_{R,r(j,0)}$ multiplets.}
\end{enumerate}

\noindent
As is clear from the above list, the existence of a unitary $\CN=2$ superconformal algebra in the IR and the resulting imposition of step 2 is the crucial part of our algorithm. Note that the role of the superpotential is indirect in step 2: it simply controls the conserved $R$-symmetry quantum numbers. In step 3, the rough form of the superpotential plays a brief but important role we will describe below.

\begin{table}[h]
 \begin{center}\renewcommand{\arraystretch}{1.2}
   \begin{tabular}{ c|c c cc ccccccccccccccc} 
 &
   $X$ & $M$ & $qq'$ & $\phi \lambda_\alpha$ 
   &  $  qq \lambda_\alpha$    &  $  q'q' \lambda_\alpha$    &  $  qq' \lambda_\alpha$ 
     & $ \lambda^2 $  & $\phi^2$ & $\phi q'q'$ & $\phi qq'$ & $\phi qq$
\\ \hline
     $r$ & $\frac35$&$ \frac65$ &  $\frac15$&  $\frac{7}{10}$&  $\frac{13}{10}$& $\frac{1}{10}$& $\frac{7}{10}$&  $1$& $\frac{2}{5}$& $-\frac{1}{5}$& $\frac{2}{5}$& 1&
     \\ \hline
     $R$ &    1& 0 & 1&$\frac12$&  $\frac32$&  $\frac32$&  $\frac32$&  1& $0$& $1$& 1& 1 &
          \\ \hline
     $j$ &   0& 0 & 0& $\frac12$&$\frac12$&$\frac12$&$\frac12$& 0 &0 &0 &0 &0
     \\ \hline
     $F^{\perp}$ &$\frac85$ & $\frac65$ & $\frac65$ & $\frac65$ & ${14\over5}$ & $\frac85$ & $\frac{11}{5}$ & 2 & $\frac25$ & $\frac45$ & $\frac75$ &2
      \end{tabular}

\hspace{.3mm} \begin{tabular}{ c|c c cc cccccccccccc} 
  & 
    $q'q' \phi \lambda_\alpha$ & $qq\phi\lambda_{\alpha}$ & $qq'\phi\lambda_{\alpha}$ &      $qq \lambda_\alpha \lambda_\beta$   &  $q'q' \lambda_\alpha \lambda_\beta$   &  $qq' \lambda_\alpha \lambda_\beta$ 
    & $\phi \lambda_\alpha \lambda_\beta$
   &  $  \lambda_\alpha \lambda_\beta \lambda_\gamma$  \\ \hline

     $r$ & $ \frac{3}{10} $  & $\frac32$ &$\frac{9}{10}$  &   $\frac{9}{5}$& $\frac{3}{5}$& $\frac{6}{5}$&  $\frac65$&  $\frac32$   
     
          \\ \hline
     $R$ &     $\frac32$ & $\frac32$& $\frac32$&  2&2&2& 1&$\frac32$ 
          \\ \hline
     $j$ &      $\frac12$ & $\frac12$ &$\frac12$ & $1$&$1$&$1$ & 1 &$\frac32$  
     \\ \hline
     $F^{\perp}$ & $\frac95$ &3& $\frac{12}{5}$ & $\frac{19}{5}$ & $\frac{13}{5}$ & $\frac{16}{5}$ & $\frac{11}{5}$ &3
 \end{tabular} 
 \caption{The result of step 1 of our algorithm: the naive UV chiral ring generators. Note that $qq':=\epsilon_{ij}q^i{q'}^j$. We use $\delta^{ab}$ to write the remaining generators involving gauge non-singlets in the first line (e.g., $qq\lambda:=\delta^{ab}(qq)_a\lambda_{\alpha b}$) and $\epsilon^{abc}$ to write the generators in the second line (e.g., $q'q'\phi\lambda_{\alpha}:=\epsilon^{abc}(q'q')_a\phi_b\lambda_{\alpha c}$). Recall that $F^{\perp}:=R+r$ is a flavor symmetry with respect to the subalgebra defined in \eqref{N1perp}.}
  \label{UVChiralRing}
\end{center}
 \end{table}

To that end, let us first implement step 1. Since any trace over $SU(2)$ generators can be written in terms of $\delta^{ab}\sim{\rm Tr}(T^aT^b)$ and $\epsilon^{abc}\sim{\rm Tr}([T^a,T^b]T^c)$, chiral ring generators will contain at most three adjoints.\footnote{Note that we do not demand that the operators obey classical relations. To understand this statement, recall that in the $\CN=1$ construction above, we have a conserved $U(1)_r$ symmetry, $r$, and a conserved $SU(2)_R$ Cartan, $R$. Consider a particular chiral operator, $\CO_1$ with $r(\CO_1)=r_1$ and $R(\CO_1)=R_1$. Since all chiral gauge-invariant operators have $R\ge0$ and  $R>0$ if $r\le0$, there are a finite number of chiral operators with the same $r$ and $R$ quantum numbers. Suppose $\hat\CO_A$ has $r(\hat\CO_A)=r_1$ and $R(\hat\CO_A)=R_1$, and suppose $\hat\CO_A$ is a product of chiral gauge invariant operators with each factor involving at most three adjoints. Let us denote $S_1$ as the set of all such $\hat\CO_A$. Then, we have the quantum operator relation
\begin{equation}\label{composite}
\CO_1=\sum_{A=1}^{|S_1|}n_{1A}\hat\CO_A+\sum_{A=1}^{|S_1|}a_{1A}(\tau,\Lambda,h_1,h_2,h_3)\hat\CO_A~,
\end{equation}
where \lq\lq$|\cdots|$" is the number of elements in the enclosed set, $h_i$ are the implicit couplings in the superpotential   \eqref{WUV}, $\tau$ is the holomorphic gauge coupling, and $\Lambda$ is the dynamical scale. Here the $n_{1A}$ are constants, while the $a_{1A}$ are functions of the couplings that vanish in the UV limit, i.e. $a_{1A}(0,0,0,0,0)=0$ (the classical chiral ring relation is of the form $\CO_1=\sum_An_{iA}\hat\CO_A$). Starting with $\CO_1$ involving four or fewer fields and working iteratively in the number of fields, we see that the chiral ring generators in the quantum theory must involve traces over either two or three fields (even though the classical chiral ring relations are modified).} Proceeding in this way results in the twenty generators in table \ref{UVChiralRing}.\footnote{At tree level, a subset of generators are  trivial in the chiral ring. These are $\phi^2,\phi q' q', \phi qq, \phi qq', qq\phi\lambda_\alpha$, and $qq'\phi\lambda_\alpha$, as one can see from contracting various fields with $\partial_X W=\partial_M W =\partial_q W=0$. Furthermore, $0=\lambda_\alpha\partial _\phi W=2 X\phi\lambda_\alpha+ M  q'q' \lambda_\alpha+   q q\lambda_\alpha$, implying  $qq\lambda_\alpha$ is not a generator in the classical chiral ring. However, in general, these classical chiral ring relations have quantum corrections.
}

Next, we implement step 2 of our algorithm. Clearly, any of the generators must either sit in an $\bar\CE$ multiplet as in \eqref{Echiral} or a $\bar\CB$ multiplet as in \eqref{Bbar}.

Let us study the $\CN=1$ chiral ring generators that can sit in $\bar\CE$ first. Any such operator has $R\le1$ and $j\le1/2$, as one can see from \eqref{Echiral}. Of the operators in table \ref{UVChiralRing}, only $X$, $M$, $qq'$, $\phi\lambda_{\alpha}$, $\lambda^2$, $\phi^2$, $\phi qq$, $\phi qq'$, and $\phi q'q'$ satisfy these constraints. Note that $\phi^2$ violates a unitarity bound and decouples.
Moreover, to be in an $\bar\CE$ multiplet, $X$, $\lambda^2$, $\phi qq$, $\phi qq'$, and $\phi q'q'$ would need to be level-two descendants of primaries with $r=8/5$, $r=2$, $r=2$, $r=7/5$, and $r=4/5$ respectively.   Since there are no such $\CN=2$ chiral operators in the MAD theory with these values of $r$ (see \eqref{N2chiral}),   we conclude that the only generators sitting in $\bar\CE$ multiplets are
\begin{equation}\label{chirgens}
M~,\ \phi\lambda_{\alpha}~,\ qq'\in\bar\CE_{6/5}~,
\end{equation}
where $M$ is the primary, $\phi\lambda_{\alpha}$ is the level-one descendant, and $qq'$ is the level-two descendant described in \eqref{Echiral} with $r=6/5$.

Next, let us discuss generators that can potentially sit in the $\bar\CB$ multiplets. The primary in \eqref{Bbar} satisfies $r>1+j$ (if $r=1+j$, we have a $\bar\CD$ multiplet, which we know is absent \cite{Buican:2021elx}), while the descendants satisfy $r>j$. We see that $q'q'\lambda_{\alpha}$, $q'q'\phi\lambda_{\alpha}$, $q'q'\lambda_{\alpha}\lambda_{\beta}$, $\lambda_{\alpha}\lambda_{\beta}\lambda_{\gamma}$, and $\phi q'q'$ do not satisfy this condition. As a result, we conclude these operators are trivial in the IR FCS.\footnote{Classically, in the chiral ring of an $\CN=1$ gauge theory, we have (e.g., see \cite{Cachazo:2002ry})
\begin{equation}
\left\{\lambda_{\alpha},\lambda_{\beta}\right\}\sim\left\{\bar Q^{\dot\alpha},[D_{\alpha\dot\alpha},\lambda_{\beta}]\right\}~,
\end{equation}
where $Q_{\alpha}=Q_{1\alpha}$, $D_{\alpha\dot\alpha}$ is the gauge covariant derivative, we define the matrix $\lambda_{\alpha}:=\lambda^a_{\alpha}\epsilon_{abc}$, and the anticommutator involves matrix multiplication. Therefore, classically, any non-trivial element of the chiral ring involving gauginos should have either spin $1/2$ or spin $0$. This statement is consistent with our entirely non-perturbative algebraic proof that $q'q'\lambda_{\alpha}\lambda_{\beta}$ and $\lambda_{\alpha}\lambda_{\beta}\lambda_{\gamma}$ are trivial in the IR chiral ring. Below, we will extend our proof to all chiral operators involving gauginos with spin different from $1/2$ or $0$.} 

The remaining operators are
\begin{eqnarray}\label{remainingLst}
&&X~,\ \lambda^2~,\ \phi qq'~,\ \phi qq~,\ qq\phi\lambda_{\alpha}~,\ qq'\phi\lambda_{\alpha}~,\ \phi\lambda_{\alpha}\lambda_{\beta}~,\ qq\lambda_{\alpha}~,\ qq'\lambda_{\alpha}~,\cr && qq\lambda_{\alpha}\lambda_{\beta}~,\ qq'\lambda_{\alpha}\lambda_{\beta}~.
\end{eqnarray}
Since $r\le1+j$, we see that these operators cannot be primaries of a $\bar\CB$ multiplet. We can then immediately see that $qq'\phi\lambda_{\alpha}$ is not a generator. Indeed, if it is a level-one descendant, then the primary has $r=7/5$, $R=1$, and $j=0$ ($j=1$ is ruled out because $r<2$). By $SU(2)_R$ and Lorentz spin, it must be of the form of a linear combination of $Mqq'$ and $(\phi\lambda_{\alpha})^2$. If this were the case, then $qq'\phi\lambda_{\alpha}\sim (qq')(\phi\lambda_{\alpha})$. If $qq'\phi\lambda_{\alpha}$ is a level-two descendant, then the primary has $R=j=1/2$ and must be of the form $M^n\phi\lambda_{\alpha}$. However, this operator is not a primary: it is a descendant of $M^{n+1}\in\bar\CE_{6(n+1)/5}$.

Let us now study the remaining generators in \eqref{remainingLst}. To that end, the primary of a $\bar \CB$ multiplet containing these operators must take the form $\CO_1 \CO_2$,
where $\CO_1$ and $\CO_2$ are (potentially) composite operators  built from \eqref{chirgens} and \eqref{remainingLst}, respectively (we do not include factors of $qq'\phi\lambda_{\alpha}$ in $\CO_2$, since it is, at best, built from operators contributing to $\CO_1$).
A crucial observation is that the charge $F^{\perp}:=R+r$ commutes with the supercharges used to relate the $\bar\CB$ members of the FCS in \eqref{Bbar} (i.e., it is a flavor symmetry of the algebra in \eqref{N1perp}). Furthermore, for operators  in \eqref{chirgens}, we have $F^{\perp}=\frac65$, while, for all the operators in \eqref{remainingLst} except for $qq'\phi\lambda_{\alpha}$, one finds $F^{\perp}\ge 7/5$ and $F^{\perp}\not\in \frac65 \mathbb Z$. Therefore, if the primary is purely made out of operators from \eqref{chirgens}---namely $\CO_2$ is trivial---then all operators in the multiplet must satisfy $F^{\perp}\in \frac65 \mathbb Z$. Since this is  in contradiction with the fact that  $F^{\perp} \not\in \frac65 \mathbb Z$ in \eqref{remainingLst},  $\CO_2$ must be non-trivial. Now, the multiplet generated by a primary of the form $\CO_1\CO_2$ (with $\CO_{1,2}\ne1$) has minimal charge  $F^{\perp}\ge 6/5+7/5= 13/5$,\footnote{Note that if $\CO_2$ has minimal $F^{\perp}$ charge, then $\CO_1$ can not be trivial,  because  $\CO_2$  is just a single operator from \eqref{remainingLst} and  thus cannot be the primary. If $\CO_2$ is composite, then $\CO_1$ may be trivial, but in that case $F^{\perp}\ge 7/5\times 2 =14/5$.}
 which rules out most operators in \eqref{remainingLst}, except for 
\be\label{remainingLst2} 
\lambda_\alpha q q~,  \qquad qq\phi\lambda_{\alpha}~, \qquad qq\lambda_{\alpha}\lambda_{\beta}~,\qquad  qq'\lambda_{\alpha}\lambda_{\beta}~.
\ee
We can repeat the same procedure to rule them out.  In particular all operators in \eqref{remainingLst2}  satisfy $3\ge F^{\perp}\ge 14/5$.  Therefore, the multiplet generated by the primary   $\CO_1\CO_2$ has $F^{\perp}\ge 14/5+6/5=4$. This logic then rules out all operators in \eqref{remainingLst2}. 

To conclude, all operators in   \eqref{remainingLst} are trivial in the chiral ring (with the possible exception of $qq'\phi\lambda_{\alpha}$ which is either trivial or else is not a generator of the FCS and satisfies $qq'\phi\lambda_{\alpha}\sim(qq')(\phi\lambda_{\alpha})$; we will rule out this latter possibility shortly). We present an alternate proof based on a case-by-case analysis in Appendix~\ref{trivialB}.
 
As an additional consistency check, note that $M$ cannot sit in a $\bar\CB$ multiplet because it has $R=0$.\footnote{Note that by definition the $\bar\CB$ multiplet has $R>0$.} Since $\phi\lambda_{\alpha}$ has $R=1/2$, it can only sit as a primary of $\bar\CB$, but $r=7/10<1+1/2$ precludes this. Finally, $qq'$ has $R=1$, so it can only sit as a primary or a level-one descendant. However, $r=1/5<1$ precludes its being a primary. If it were a level-one descendant, the primary would have to be $\phi\lambda_{\alpha}$ on $U(1)_r$ grounds. Fortunately, this scenario is already ruled out.

Therefore, to summarize, we learn that the IR FCS is generated by the three operators of \eqref{chirgens} living in the $\bar\CE_{6/5}$ multiplet we expect from the Seiberg-Witten solution. Moreover, all the other putative generators in table \ref{UVChiralRing} are in fact trivial in the FCS or correspond to products of the generators. 

Now we would like to understand if we can construct any $\bar\CB$ multiplets (i.e., step 3 of our algorithm). To that end, we would need to build the corresponding chiral operators out of $M$, $\phi\lambda_{\alpha}$, and $qq'$. We can immediately see that constructing an operator of the form $\bar\CB_{1/2,19/10(1/2,0)}$ is impossible: the only way to construct the primary is via $M\phi\lambda_{\alpha}$, but this is clearly a level-one descendant of $\bar\CE_{12/5}$.\footnote{We know from the OPE selection rules in \cite{Beem:2014zpa} that $\bar\CB_{1/2,19/10(1/2,0)}$ cannot appear in the OPE of two identical $\bar\CE_r$ multiplets. Note, however, that if we have multiple $\bar\CE_r$ multiplets, such operators can appear in the OPE.}

More generally, it is useful to find constraints among the operators in \eqref{chirgens}. For example, from the discussion in \cite{Komargodski:2020ved}, we expect both $\phi\lambda_{\alpha}$ and $qq'$ to be nilpotent. Of course, $\phi\lambda_{\alpha}$ is nilpotent since it is a fermion. Using \eqref{WUV}, it is easy to see that $qq'$ is as well. Indeed, classically in the $\CN=1$ chiral ring, we have
\begin{equation}\label{qqprnil}
0=\partial_{\phi^a}W\cdot (q'q')^a=(2X\phi^a+M(q'q')^a+(qq)^a)(q'q')^a=(qq')^2~,
\end{equation}
where we have used our above analysis to show that $X$ vanishes in the IR FCS.

We can ask if \eqref{qqprnil} can be modified in the quantum theory. By $U(1)_r$ conservation, any putative quantum corrections would have $r=2/5$. Our analysis of $\CN=2$ superconformal representation theory and its consequences are fully non-perturbative in nature and so the operators we ruled out cannot appear. Moreover, any correction cannot involve $M$ or $\phi\lambda_{\alpha}$ either since there is no negative $U(1)_r$ charged chiral operator to compensate for their larger $U(1)_r$ charge. As a result, we see that \eqref{qqprnil} is exact. In particular, we see that $qq'$ is minimally nilpotent in the quantum theory (it vanishes at quadratic order).

Another useful constraint is that
\begin{equation}\label{lqqpr}
(\phi\lambda_{\alpha})(qq')=0~,
\end{equation}
in the IR FCS. To understand this statement, note that this operator has $R=3/2$ and $r=9/10$. Since $R>1$, it can only sit in a $\bar\CB$ multiplet. Clearly, it cannot be a primary or a level-one descendant since $\left[Q^{1\alpha},((\phi\lambda_{\alpha})(qq'))\right]=(qq')^2=0$ in the chiral ring. If it is a level-two descendant, then the primary has $r=19/10$ and $R=1/2$, which means it is of the form $M\phi\lambda_{\alpha}$ (but this is a level-one descendant of $\bar\CE_{12/5}$).

Now we can ask if it is at all possible to construct a primary of a $\bar\CB$ multiplet. Note that our reasoning above implies
\begin{equation}
\left\{Q^{1\alpha},\left[Q^1_{\alpha},M^n(\phi\lambda_{\alpha})^2\right]\right\}=0~,\ \ \ \left\{Q^{1\alpha},\left[Q^1_{\alpha},M^nqq'\right]\right\}=0~,
\end{equation}
in the FCS. As a result, $M^n(\phi\lambda_{\alpha})^2$ and $M^nqq'$ cannot be primaries of a $\bar\CB$ multiplet. To construct such a primary, we must therefore use exactly one $\phi\lambda_{\alpha}$. However, by \eqref{lqqpr} such a primary must not involve $qq'$. The only option remaining is $M^n\phi\lambda_{\alpha}$, but this is a level-one descendant of an $\bar\CE$ multiplet.\footnote{In fact, one can easily see that $(\phi\lambda_{\alpha})^2$ and $Mqq'$ have the same quantum numbers $R=1$ and $r=7/5$. Therefore, we may expect a relation between them
\be\label{CRAB}
a(\phi\lambda_{\alpha})^2+b Mqq'=0~,
\ee
where $a,b\in\mathbb{C}$ are constants. Such a relation must exist because they can only appear in the level-two $Q^1_\alpha$ descendant of $M^2$. This logic implies $(\phi\lambda_{\alpha})^2$ and $Mqq'$  cannot be two independent operators in the chiral ring. It would be interesting to determine the coefficients $a$ and $b$. This computation is beyond the scope of this paper.
}

We have therefore established Claim 1 and the attendant Corollaries 1 and 2. In particular, Corollary 2 implies that, as promised, there are no $\bar\CB$ or $\CB$ multiplets in the MAD theory.

In the next section we will use the absence of these multiplets, combined with the superconformal index, to make exact statements about the multiplicities of $\bar\CC_{0,r(j,0)}$ (and $\CC_{0,-r(0,j)}$) multiplets for all possible $r$ and $j$.

\newsec{Exact results on the semi-short spectrum: beyond the Schur sector}\label{semiShort}
Combining the results of the last section with those of \cite{Buican:2021elx}, let us summarize the known part of the MAD short multiplet spectrum. First, we list the number of multiplets containing FCS operators but no Schur operators\footnote{Multiplicities of conjugate multiplets are fixed by CPT to be equal to those we list.}
\begin{equation}\label{NEB}
N_{\bar\CE_r}=\delta_{r,6n/5}~, \;(n \in \mathbb Z_+), \qquad\ \ \ N_{\bar\CB_{R, r(j,0)}}=0~.
\end{equation}
Next, let us list the number of multiplets containing operators in both the FCS and the Schur sector
\begin{equation}
N_{\hat B_R}=N_{\bar\CD_{R(j,0)}}=0~.
\end{equation}
Finally, let us list the number of   multiplets containing operators only in the Schur sector. The number of  $\hat\CC$  multiplets is given by the following generating function \cite{Buican:2021elx}
 \be\label{NChat}
\frac{ x^{R(R+2)} }{(1-x^2) (1-x^3) \cdots (1-x^{R+1})}=\sum_{ 2j=0}^\infty x^{2j} N_{\hat\CC_{R(j,j)}}~. 
 \ee

In order to fully determine the short multiplet spectrum of the MAD theory, all that remains is to fix the multiplicity of $\bar\CC_{R,r(j,\bar j)}$ multiplets. Finding the full spectrum of these multiplets is an involved problem that we will return to \cite{BBJ2}. Interestingly, we will see in later sections that these multiplets act as a kind of \lq\lq glue" that link the Schur and chiral sectors (we will make this notion more precise via a conjecture).

In this section, we will argue that it is possible to determine the exact spectrum of $\bar\CC_{0,r(j,0)}$ multiplets (and we will interpret this fact physically in section \ref{Minimality}). In particular we claim that
\begin{equation}\label{Cclaim}
N_{\bar\CC_{0,r(j,0)}}=\begin{cases}
1~, &\text{if}\ r={6\over5}n+{j\over5}~,\ j=0,1~,\ n\in\mathbb{Z}_{>0}~,\\
0~, &\text{otherwise}~.
\end{cases}
\end{equation}

Our main tool for deriving \eqref{Cclaim} is the superconformal index of the MAD theory \cite{Maruyoshi:2016tqk}. However, we must first overcome certain ambiguities in the index in order to find the precise spectrum of our multiplets of interest. To understand this point, let us first recall the definition of the superconformal index  %
\be\label{Indyutau}
  \cI  ={\rm Tr} (-1)^F p^{j +\bar j +r} q^{\bar j -j+r}  t^{R-r}~,
  \ee
where the trace is over the Hilbert space of local operators, $(-1)^F$ is the fermion number, $j,\bar j$ are the left and right Lorentz spins, $r$ is the $U(1)_r$ charge,  and $R$ is the $SU(2)_R$ weight introduced before. The superconformal index counts the contributions of short multiplets up to recombination (and long multiplets do not contribute to the index). Below, it turns out to be more convenient to use another set of fugacities $u,y,\tau$ which are related to $p,q,t$ through
 \be \label{yutau2}
 p=\tau^3 y~, \qquad q=\tau^3/y~,   \qquad u =pq/t~.
 \ee
A big advantage of working with $u,y,\tau$ is that we can study the index perturbatively in  $\tau$ and exactly in $u$ and $y$.

The index of the SCFT is given by the sum of contributions from  various short multiplets. For the MAD theory, the only short multiplets are of type $\hat\CC$, $\bar\CE$, and $\bar \CC$ (the conjugate $\CE $ and $\CC$ multiplets do not contribute to the index). Therefore, we can write the index as
 \be\label{Indexdecomp}
 \CI_\text{MAD}=\sum_\Xi N_\Xi \CI_\Xi=1+ \CI_{\hat \CC}+ \CI_{\bar\CE}+ \CI_{\bar\CC}~, \qquad N_\Xi\in\mathbb N~,
 \ee
 where the leading contribution is from the vacuum.   In order to furnish the decomposition,  let us first recall the index contributions from the individual $\bar \CC, \hat \CC$, and $\bar\CE$ multiplets of interest
\beqn
{\cal I}_{{\hat \CC}_{R(j, \bar j)}}&=&
(-1)^{2(j+ \bar j)}p^{j}q^{j}t^{R-j+\bar j-1}\frac{t-p q}{(1-p)(1-q)}
\left[p^{\frac12} q^{\frac12} t\chi_{j+{\frac12}}\left(\sqrt{\frac{p}{q}}\right)
-pq\chi_{j}\left(\sqrt{\frac{p}{q}}\right)\right]~,\nonumber\\
{\cal I}_{\bar\CE_{r }}&=&  p^{r-1}q^{r-1}t^{-r} \frac{(t-p)(t-q)}{(1-p)(1-q)}\ \label{indexE}~,
\nonumber\\
{\cal I}_{\bar\CC_{R,r(j, \bar j)}}&=&
-(-1)^{2(j+ \bar j)}p^{\bar j+r}q^{\bar j+r}t^{R-r-1}\frac{(t-pq)(t-p)(t-q)}{(1-p)(1-q)}\chi_{j}\Big(\sqrt{\frac{p}{q}}\Big)~, 
\label{Cbarindex}
\eeqn
where $ \chi_{j} (x):=\big(x^{2j+1}-x^{-{(2j+1)}}\big)\big/\big(x-x^{-1}\big)$ is the character of the spin-$j$ representation of $SU(2)$.

Our goal here is to decompose the superconformal index into the contributions from various multiplets, $\Xi$, and thus compute the multiplicities, $N_\Xi$. For    ${\bar\CE}$ and ${\hat\CC} $  multiplets, the multiplicities can be read off unambiguously from the Coulomb branch and Macdonald indices respectively\footnote{These are special limits of the superconformal index \cite{Gadde:2011uv}.} (the latter statement follows non-trivially from the considerations in \cite{Buican:2021elx}). The results are given in \eqref{NEB} and \eqref{NChat}. The next step is to find the multiplicities of the $\bar \CC$ multiplets. Before setting up the computation, let us first remark on a few subtleties.

In general, given \eqref{Indyutau} and some multiplet of type $\Xi$, there are several potential obstacles to finding the precise number of such multiplets, $N_{\Xi}$:
\begin{enumerate}
\item{Index cancellations due to the fact that there are multiplets, $\Xi'_i$, that can combine with $\Xi$ to form a long multiplet. In this case, even if the short multiplets have not recombined to a long multiplet (i.e., the long multiplet is at its unitarity bound), the index contribution of $\Xi$ is canceled.
\item Cancelling index contributions can also be generated by multiplets that mimic the index contributions of the $\Xi'_i$ (even if they cannot recombine with $\Xi$ to form a long multiplet).}
\item{The contribution of $\Xi$ may potentially be the same as the contribution of a multiplet with different quantum numbers.}
\item{In practice, we need to expand the index in powers of fugacities, but it may be difficult to disentangle the leading contribution from $\Xi$ and subleading contributions from other multiplets.}
\end{enumerate}

Let us analyse the case of interest, $\Xi=\bar\CC_{0,r(j,0)}$. First, note that our results in the previous section rule out obstacle 1. Indeed, such multiplets can only recombine as follows (e.g., see \cite{Beem:2013sza})
\begin{equation}\label{ACB}
\CA^{r+2}_{0,r(j,0)}=\bar\CC_{0,r(j,0)}\oplus\bar\CB_{1,r+1(j,0)}~,
\end{equation}
where $\CA$ is a long multiplet. However, we have shown there are no $\bar\CB$ multiplets.\footnote{Note that the $\bar\CB_{R,r(j,0)}$ multiplet contributes to the index in the same way as a would-be $\bar\CC_{R-1/2,r-1/2(j,-1/2)}$ multiplet. This fact explains the index cancelation that results from the recombination in \eqref{ACB}.}
  
Let us now consider the remaining obstacles 2-4 simultaneously. As discussed above, since we know the multiplicity of all $\bar\CE$ and $\hat\CC$ multiplets, we can subtract their contributions to the index. Therefore, we need only consider to what extent other $\bar\CC$ multiplets can lead to obstacles 2-4. To see these observables do not pose a problem, consider the leading-order in $\tau$ index contribution from a $\bar\CC_{R,r(j,\bar j)}$ multiplet \eqref{Cbarindex}
\be\label{leadingCC}
\CI_{\bar\CC_{R,r(j,\bar j)}}= (-1)^{2 j+2 \bar j+1} \tau ^{6 (\bar j+R+1)}(1-u) u^{r-R}\chi_{j}(y)+\cdots~.
\ee
Since $\bar j, R\ge0$, we see that any $\bar\CC_{R,r(j,\bar j)}$ contributions with $R\ne0$ or $j\ne0$ will be subleading in $\tau$ compared to the $\bar\CC_{0,r(j,0)}$ index contributions
\be\label{CCcont}
\CI_{\bar\CC_{0,r(j,0)}}= (-1)^{2 j+1} \tau ^{6}(1-u) u^{r}\chi_{j}(y)+\cdots~.
\ee
 
Moreover, these contributions clearly distinguish different $r$ and $j$. As a result, we see that we can use the index to unambiguously extract the multiplicities of the $\bar\CC_{0,r(j,0)}$ multiplets.

Let us now proceed to extract this unambiguous data. To that end, the index of the MAD theory is given by \cite{Maruyoshi:2016tqk}
 \be\label{index2}
 \cI_\text{MAD}=\kappa \frac{\Gamma_e\Big( u^{\frac65}\Big) }{\Gamma_e\Big( u^{\frac25}\Big)}
 \oint_C \frac{dz}{2\pi i z} \frac{ \Gamma_e\Big(  z^{\pm  }\tau^3u^{-{1\over10}}\Big) 
\Gamma_e\Big(  z^{\pm  }\tau^3u^{-{7\over10}}\Big) 
 \Gamma_e\Big(  z^{\pm 2,0}u^{\frac15} \Big)}
 {2\Gamma_e (z^{\pm 2} )}~,
 \ee
 where we use the notation $f(z^\pm) :=f(z) f(z^{-1})$,  $f(z^{\pm 2, 0}):=f(z^2)f(z^{-2}) f(z^0)$, the elliptic Gamma function is defined as
 \be
 \Gamma_e(x) 
 :=  \Gamma_e(x; p,q)
 =\prod_{m,n=0}^\infty \frac{1-x^{-1} \tau^{3(m+n+2)}y^{m-n}}{1- x\tau^{3(m+n)}y^{m-n}}~,
 \ee
 and
 \be
 \kappa=(\tau^3y;\tau^3y)(\tau^3y^{-1};\tau^3y^{-1})~, \qquad (z;x):=\prod_{n=0}^\infty (1-zx^n)~.
 \ee
Up to the overall order in \eqref{CCcont}, we need to expand the elliptic gamma function as follows\footnote{Here we expand to order beyond $\tau^6$, because the    $ \tau^{ 3}$  factor in the argument of $\Gamma_e$  \eqref{index2} may reduce the power of $\tau$ in $\Gamma_e$.}
  \bea \nonumber
 \Gamma_e (x)&\to&
\frac{1}{1-x}
    +\frac{  \left(y^2+1\right) x}{y(1- x)}\tau ^3
     +\frac{  \left(-\left(\left(y^4+y^2+1\right) x^3\right)-\left(y^4+y^2+1\right) x^2+y^2\right)}{y^2 (x-1) x}\tau ^6
      \\&&
       -\frac{ \left(y^2+1\right) \left(y^4 x^2 \left(x^2+x+1\right)+y^2 \left(x^3-x-1\right)+x^2 \left(x^2+x+1\right)\right)}{y^3 (x-1) x}\tau ^9 
\label{gammae} \\&&  
+\frac{1}{y^4 (x-1) x}\Big (-\left(y^8+1\right) x^2+\left(2 y^4+3 y^2+2\right) y^2 x+y^6+y^4+y^2  \nonumber\\&&-(1 + y^4) (2 + 3 y^2 + 2 y^4) x^3 
 - (1 + y^2 + y^4)^2 x^4 - (1 + 
      y^2 + y^4 + y^6 + y^8) x^5\Big) \tau^{12}+\cdots~.
      \nonumber
 \eea
 Then the index can be written as
 \be
 \cI_\text{MAD} =\oint \frac{dz}{2\pi i z}  \Big( A_0 +A_1 \tau^3+ A_2 \tau^6+\cdots\Big)~,
 \ee
 where 
 \be
 A_0=\frac{\left(1-u^{2/5}\right) \left(1-\frac{1}{z^2}\right) \left(1-z^2\right)}{2 \left(1-u^{1/5}\right) \left(1-u^{6/5}\right) \left(1-\frac{u^{1/5}}{z^2}\right) \left(1-u^{1/5} z^2\right)}~.
 \ee
Note that $A_1$ and $A_2$ can also be obtained explicitly, but we do not write down the corresponding complicated expressions  here. Using the residue theorem at the $z=0, \pm u^{\frac{1}{10}}$ poles, we can evaluate the integral and find
 \bea\label{MADindex}
 \cI_\text{MAD}&=&
 \frac{1}{1-u^{6/5}}
 - \frac{  (1-u)  {u^{1/5}}}{1-u^{6/5} }\chi_{\frac12}(y) \tau ^3
 \\&&
  +\frac{  \left(1+u^{2/5}  \right) (1-u)
   \Big(1 - u^{2/5} -u  ( 1 - u^{2/5} + u^{4/5})  \chi_1(y)\,\Big)}{u^{4/5} \left(1-u^{6/5}\right)}\tau ^6
  +\CO(\tau^7)~.
  \nonumber
 \eea

As discussed above, in order to solve for $N_{\CC_{0,r(j,0)}}$, we must subtract the contributions from $\hat\CC$ and $\bar\CE$ multiplets in \eqref{NChat} and \eqref{NEB} respectively. To that end, the $\hat\CC$ index contributions take the form
 \be
 \cI_{\hat \CC}  =\sum_{R,j} N_{\hat\CC_{R(j,j)}} 
 \CI_{\hat \CC_{R(j,j)}}  =\tau ^6 (u-1) +\frac{ (u-1)^2 \left(y^2+1\right)}{u y}\tau ^9 +\CO(\tau^{10})~,
 \ee
while the $\bar\CE$ contributions take the form
\bea
 \cI_{\bar\CE}&=&\nonumber
 \sum_{k=1}^\infty  \CI_{\bar\CE_{\frac65k}}
= \frac{u^2 y-\tau ^3 u \left(y^2+1\right)+\tau ^6 y}{u^{4/5} \left(u^{6/5}-1\right) \left(y-\tau ^3\right) \left(\tau ^3 y-1\right)}
 \\&=&
  \frac{ u^{6/5}}{1-u^{6/5}}
 - \frac{  (1-u)  {u^{1/5}}}{1-u^{6/5} }\chi_{\frac12}(y) \tau ^3
 +\frac{  (1-u) (1- u\chi_1(y))}{  \left(1-u^{6/5}  \right)u^{4/5}}\tau ^6 
 +\CO(\tau^{7})~.
\eea
Subtracting  these contributions (along with the contribution of the identity operator, as summarized in \eqref{Indexdecomp}) from the index \eqref{MADindex} yields the $\bar\CC$ contribution
  \bea\label{ICbar}
 \cI_{\bar\CC} &=&\nonumber  -\frac{  (1-u) u^{6/5}  }{1-u^{6/5}}\tau ^6
 -\frac{ (1-u) u^{7/5} \chi_1(y)\,   }{1-u^{6/5}}\tau ^6+\CO(\tau^7)
\\&=&
 \sum_{k=1}^\infty \Big(  - \tau ^6(1-u) u^{6/5k} 
 - \tau ^6(1-u) u^{6/5k+1/5}\chi_1(y)   \Big) +\CO(\tau^7)~.
 \eea
Comparing this expression with the leading term in the index of the $\bar\CC$ multiplet \eqref{leadingCC}, we can then easily figure out the corresponding multiplet.  Indeed, the spectrum of $\bar\CC_{0,r(j,0)}$ operators is given by
\begin{equation}\label{Cclaim2}
N_{\bar\CC_{0,r(j,0)}}=\begin{cases}
1~, &\text{if}\ r={6\over5}n+{j\over5}~,\ j=0,1~,\ n\in\mathbb{Z}_{>0}~,\\
0~, &\text{otherwise}~.
\end{cases}
\end{equation}
Therefore, we have arrived at our main claim!  As we will discuss later, these multiplets appear in OPEs involving only $\bar\CE_{\frac65n}$ and $\hat \CC_{0(0,0)}$ multiplets.\footnote{One can proceed further and find that, at the next order
 \beqn
  \cI_{\bar\CC} &=&  -\frac{  (1-u) u^{6/5} \tau ^6 }{1-u^{6/5}}  \Big( 1+\frac{u-1}{u} \chi_{\frac12}(y)\, \tau^3\Big) 
 -\frac{ (1-u) u^{7/5} \tau ^6     }{1-u^{6/5}}  \Big( \chi_1(y) 
 +\frac{u-1}{u} \big(\chi_{\frac12}(y)+\chi_{\frac32}(y) \big) \, \tau^3 \Big)
 \nonumber\\&&
  -\frac{  (1-u) u^{6/5} \tau ^9 }{1-u^{6/5}} \chi_{\frac12}(y)
 -\frac{ (1-u) u^{13/5}\tau ^9   }{1-u^{6/5}} \chi_{\frac32}(y)
 + \CO(\tau^{10})~.
 \eeqn
 It is easy to check that the first line is the contribution from the multiplets in \eqref{Cclaim2}, while the second line comes from $ {\bar\CC_{0,r(j,\frac12)}}$ or ${\bar\CC_{1/2,r+1/2(j,0)}}$ multiplets whose multiplicities are subject to the condition
  \be\label{Cbarcontri}
 N_{ {\bar\CC_{0,r(j,\frac12)}}}- N_{ {\bar\CC_{ \frac12 ,r+\frac12(j,0)}}}=
 \begin{cases}
 1, & \quad \text{if } \;
 r=\frac{6}{5}n +\frac{7 }{5}(j-\frac12), \;\quad j=\frac12,\,\frac32 , \qquad n\in\mathbb Z_{>0}~,
 \\
 0,& \quad \text{otherwise}~.
 \end{cases}
 \ee 
  Purely at the level of the index, we are unable to determine the multiplicities unambiguously due to the relation $\CI_{\bar\CC_{R,r(j,\bar j)}}+\CI_{\bar\CC_{R+\frac12,r+\frac12(j,\bar j-\frac12)}}=0$. It would be interesting to resolve these ambiguities using further physical input. However, even at the above level of precision, we will soon see that \eqref{Cbarcontri} is consistent with the quantum numbers of operators appearing in OPEs involving only $\bar\CE_{{6n\over5}}$ and $\hat\CC_{0(0,0)}$ multiplets.}

Before concluding this section, let us note that \eqref{ICbar} also provides a check of our results in the previous section since we can immediately conclude that there are no $\bar\CB_{1/2,r(j,0)}$ multiplets. Indeed, general $\bar\CB$ multiplets contribute as
\begin{equation}\label{IBhalf}
\CI_{\bar\CB_{R,r(j,0)}}=\CI_{\bar\CC_{R-\frac12,r-\frac12(j,-\frac12)}}= (-1)^{2 j } \tau ^{6R}(1-u) u^{r-R}\chi_{j}(y)+\cdots~.
\end{equation}
Therefore, we see that multiplets of the form $\bar\CB_{1/2,r(j,0)}$ are unambiguously captured by the index at $\CO(\tau^3)$ once the $\bar\CE$ multiplets have been subtracted (this statement is related to the fact that these multiplets do not recombine). Indeed, since there is no $\CO(\tau^3)$ term in \eqref{ICbar}, these multiplets cannot be present.

In the next section, we will compare \eqref{Cclaim2} with the theory at generic points on the MAD Coulomb branch (i.e., the free $\CN=2$ Abelian gauge theory). This comparison will build intuition that we will use in the subsequent section to understand the universality of the spectrum of operators we are discussing.

\newsec{Comparison with the free vector}\label{FrVec}
In \cite{Buican:2021elx}, we saw that the MAD theory shares an infinite amount of Schur sector OPE data with the free $\CN=2$ vector multiplet. In this section, we will extend these observations to the full chiral sector (see section \ref{chirRing}) and the part of the semi-short spectrum described in section \ref{semiShort}.

\subsec{The full chiral sector}
Let us first consider the FCS. Any such operator takes the form $\phi^n$, $\phi^n\lambda^1_{\alpha}$, or $\phi^n\lambda^1\lambda^1$ for any $n\ge1$ and $\alpha=1,2$.\footnote{Here $\lambda^1_{\alpha}$ is the $SU(2)_R$ highest-weight component of the gaugino doublet. The fact that the chiral operator is the highest-weight component follows from the general discussion in Appendix \ref{chiralOPgeneral}.} Clearly, the first type of operator is a primary of $\bar\CE_n$ (if $n>1$; otherwise, $\phi$ is the primary of a $\bar\CD_{0(0,0)}$ multiplet). The second and third types of operators are level one and two descendants of $\bar\CE_n$ (if $n>1$; otherwise, they are level one and two descendants of a $\bar\CD_{0(0,0)}$ multiplet). Therefore, just as in the MAD case, there are no $\hat\CB\oplus\bar\CB\oplus\CB$ multiplets:
\begin{equation}
\left(\hat\CB\oplus\bar\CB\oplus\CB\right)_{\rm Free Vect.}=\left(\hat\CB\oplus\bar\CB\oplus\CB\right)_\text{MAD}=\emptyset~.
\end{equation}

It is then easy to see that, up to a $U(1)_r$ rescaling and the additional equations of motion in the $\bar\CD_{0(0,0)}$ multiplet, the FCS of the MAD theory and the FCS of the free vector are isospectral. In particular, equations of motion project out the $R=1$ states in $\bar D_{0(0,0)}$ (i.e., we set the $D$ and $F$ auxiliary fields to zero) and constrain other components of the multiplet.

\subsec{The semi-short sector}
Next, consider the semi-short $\bar\CC_{0,r(j,0)}\oplus\CC_{0,-r(0,j)}$ spectrum. As in the MAD case, the absence of $\bar\CB$ multiplets means that the index unambiguously computes the multiplicities of $\bar\CC_{0,r(j,0)}$ multiplets.

To begin, let us write the free vector index in terms of the $u,\tau, y$ variables of \eqref{yutau2}
\begin{equation}
\cI_{\rm Free Vect.}=\left[\prod_{k\ge1}(1-\tau^{3k}y^k)(1-\tau^{3k}y^{-k})\right]\times\prod_{\ell,m\ge0}{1-u^{-1}\tau^{3(\ell+m+2)}y^{\ell-m}\over1-u\tau^{3(\ell+m)}y^{\ell-m}}~.
\end{equation}
As in our analysis of the MAD theory, we should focus on the term at order $\tau^6$. Recall that the leading-order contribution in $\tau$ from a $\hat\CC_{R(j,\bar j)}$ multiplet is $\tau^{6(1+\bar j+R)}$. As a result, to capture the contributions from the semi-short multiplets in question, we should subtract the contribution from the stress tensor multiplet and the $\hat\CC_{0(1,0)}$ multiplet (with Schur operator $\lambda_+^1\partial_{+\dot+}\lambda^1_+$)
\begin{equation}\label{C010}
\cI_{\hat\CC}=(u-1)(1+u\chi_1(y))\tau^6+\cdots~.
\end{equation}
We should also subtract the contribution of the $\bar\CD_{0(0,0)}\oplus\CD_{0(0,0)}$ multiplet (all other Schur multiplets are of type $\hat\CC_{R(j,\bar j)}$ \cite{Buican:2021elx})
\begin{equation}
\cI_{\bar\CD\oplus\CD}=u+(u-1)\chi_{1/2}(y)\tau^3+(-u^{-1}+u\chi_1(y)+1-\chi_1(y))\tau^6+\cdots~.
\end{equation}
Finally, we should subtract the contributions from the Coulomb branch operators, $\bar\CE_n$ (with $n\ge2$)
\begin{eqnarray}
\cI_{\bar\CE}&=&{u^2\over 1-u}+{u-1\over1-u}u\chi_{1/2}(y)\tau^3+{u-1\over1-u}(-1+u\chi_1(y))\tau^6+\cdots\cr&=&{u^2\over 1-u}-u\chi_{1/2}(y)\tau^3+(1-u\chi_1(y))\tau^6+\cdots~.
\end{eqnarray}
Therefore, the contributions involving $\bar\CC$ are
\begin{equation}
\cI_{\bar\CC}=-u(1+u\chi_1(y))\tau^6+\cdots~.
\end{equation}
Note that the $y$-independent term corresponds to
\begin{equation}
\sum_{n\ge1} \CI_{\bar\CC_{0,n(0,0)}}=-\sum_{n\ge1}(1-u)u^n\tau^6=-u\tau^6~,
\end{equation}
while the $y$-dependent term corresponds to
\begin{equation}
\sum_{n\ge2} \CI_{\bar\CC_{0,n(1,0)}}=-\sum_{n\ge2}(1-u)u^n\chi_1(y)\tau^6=-u^2\chi_1(y)\tau^6~.
\end{equation}
In particular, we see that the $\bar\CC_{0,r(j,0)}\oplus\CC_{0,-r(0,j)}$ spectrum of the MAD theory and the free vector are in one-to-one correspondence for $j=0$. Note that for $j=1$, $\bar\CC_{0,1,(1,0)}$ hits a unitarity bound and is actually a $\hat\CC_{0(1,0)}$ multiplet (we therefore included it in \eqref{C010}). However, it is natural to include this multiplet in the map between $\bar\CC\oplus\CC$ sectors of the MAD theory and the free vector. Indeed, we then get (as in the case of the FCS) a simple one-to-one map, up to the vanishing of a finite number of states (the additional null states in $\hat\CC_{0(1,0)}$). Moreover, we expect MAD $\bar\CC\oplus\CC$ multiplets to be sources of $\hat\CC$ multiplets in the IR (since $U(1)_r$ is broken).

Let us examine these operators more carefully. To that end, the $\bar\CC_{0,n(0,0)}$ multiplet has a primary arising from the normal-ordered product of a chiral operator with the stress tensor multiplet primary
\begin{equation}\label{C0}
\bar\CC_{0,n(0,0)}\supset\phi^{n}\times\phi\phi^{\dagger}=\phi^{n+1}\phi^{\dagger}~,
\end{equation}
where $\phi^n$ is the $\bar\CE_n$ primary ($\bar\CD_{0(0,0)}$ if $n=1$), and $\phi^{\dagger}\phi$ is the dimension two primary of $\hat\CC_{0(0,0)}$. On the other hand, the $\bar\CC_{0,n(1,0)}$ multiplets take the form
\begin{equation}\label{Chalf}
\bar\CC_{0,n(1,0)}\supset\epsilon_{ij}\lambda^i_{\alpha}\lambda^j_{\beta}\phi^{n-1}+\gamma F_{\alpha\beta}\phi^n~,
\end{equation}
where the coefficient $\gamma$ can be fixed by demanding that the above operator is annihilated by the $S^{\ell}_{\delta}$ supercharges. Up to a shift by a descendant, \eqref{Chalf} appears in the normal-ordered product of $\phi^m$ and the level-two $\bar\CE_{n-m+1}$ descendant, $\epsilon_{ij}\left\{Q^i_{\alpha},\left[Q^j_{\beta},\phi^{n-m+1}\right]\right\}$.

As a result, we see that there is a simple map between the set of MAD $\bar\CC_{0,r(j,0)}$ multiplets and those of the free $U(1)$ theory.\footnote{At the special points on the MAD moduli space where free hypers appear, the map is still simple: the role of the stress tensor is played by the sum of the free vector and free hyper stress tensors.} Indeed, consider the $j=0$ multiplets first (they cannot mix with the $j=1/2$ multiplets under RG flow). In the case of the MAD theory, the minimal $U(1)_r $ charge of a $\bar\CC_{0,r(0,0)}$ primary is $r_\text{min}=6/5$, while it is $r_\text{min}=1$ in the case of the free vector. The $U(1)_r$ charge quantization (the difference in $r$ between successive $\bar\CC_{0,r(0,0)}$ multiplets) is $\delta r=6/5$ for the MAD theory and $\delta r=1$ for the free vector. Therefore, we need only apply a $r\to {5\over6}r$ rescaling in order to find a map between the $j=0$ sectors. Consider now the $j=1$ sector. Here, $r_{\rm min}=2\times 6/5-1=7/5$ in the MAD theory and $r_{\rm min}=2\times 1-1=1$ in the free vector theory (where we include $\hat\CC_{0(1,0)}$ in this discussion; otherwise, $r_{\rm min}=2$). The charge quantization is as before ($\delta r=6/5$ for MAD and $\delta r=1$ for the free vector). Therefore, to find a map between sectors, we need to rescale $r_{\rm min}\to {5\over7}r_{\rm min}$ and $\delta r\to{5\over6}\delta r$. Since $U(1)_r$ is broken in any flow to the free vector multiplet, we expect these mappings to be mappings of sectors rather than of individual operators (i.e., there will be mixing). We comment on these ambiguities in the next section.

\newsec{Minimality of the spectrum}\label{Minimality}
The quantum numbers in our result on the spectrum of $\bar\CC_{0,r(j,0)}$ multiplets \eqref{Cclaim2} suggest that
\begin{equation}
\bar\CE_{6n/5}\times\hat\CC_{0(0,0)}\supset \bar\CC_{0,6n/5(0,0)}~,\ \ \ \bar\CE_{6n/5}\times\bar\CE_{6n'/5}\supset\bar\CC_{0,6(n+n')/5-1(1,0)}~,
\end{equation}
with non-vanishing OPE coefficients. At the level of components, we expect the following normal-ordered products appearing in the OPEs below are non-vanishing
\begin{eqnarray}\label{MJ}
M^n(x)\times J(0)&=&\cdots+ M^nJ(0)+\cdots~, \cr M^n(x)\times
 \epsilon_{ij}\left\{Q^i_{\alpha},\left[Q^j_{\beta},  M^{n'}\right]\right\}(0)&=&\cdots+M^n
 \epsilon_{ij}\left\{Q^i_{\alpha},\left[Q^j_{\beta}, M^{n'}\right]\right\}(0)+\cdots~,
\end{eqnarray}
where $M$ is the $\bar\CE_{6/5}$ $\CN=2$ chiral primary introduced previously, and $J$ is the primary of $\hat\CC_{0(0,0)}$. In the case of the free vector $M\to\phi$, $J\to\phi^{\dagger}\phi$, and the corresponding non-vanishing normal-ordered products are given by \eqref{C0} and \eqref{Chalf} respectively (where the latter equation holds in the OPE up to mixing with descendants).

More generally, we expect that in a unitary $\CN=2$ SCFT with a Coulomb branch, the following normal-ordered products are non-zero
\begin{eqnarray}
\sO(x)\times J(0)&=&\cdots+\sO J(0)+\cdots~,\cr\ \ \sO(x)\times \epsilon_{ij}\left\{Q^i_{\alpha},\left[Q^j_{\beta},\sO'\right]\right\}(0)&=&\cdots+\sO\epsilon_{ij}\left\{Q^i_{\alpha},\left[Q^j_{\beta},\sO'\right]\right\}(0)+\cdots~,
\end{eqnarray}
where $\sO$ and $\sO'$ are non-trivial $\bar\CE$ $\CN=2$ chiral primaries   that can acquire non-vanishing vevs, $\langle\sO\rangle,\langle\sO'\rangle\ne0$. The main reason we expect the above normal-ordered products to not vanish is that the entire $\hat\CC_{0(0,0)}$ multiplet can be unambiguously tracked along any $\CN=2$-preserving RG flow since it contains the stress tensor \cite{Antoniadis:2010nj,Cordova:2018acb} (see also \cite{Abel:2011wv} and \cite{Bertolini:2021cew} for 4D $\CN=1$ and 5D $\CN=1$ discussions respectively). Since $\sO$ and $\sO'$ can also be tracked along an RG flow to the Coulomb branch,\footnote{This statement holds when there is a Seiberg-Witten description or some generalization thereof that allows one to compute the mixing; see \cite{Argyres:2015gha} for some examples.} we can track the above OPEs to the Coulomb branch where they are guaranteed to be non-vanishing as in the single vector multiplet case discussed in the previous section. In particular, the normal-ordered product does not vanish in the IR.
Since the non-vanishing normal-ordered product in the IR must come from a non-trivial operator in the UV, we arrive at the following claim\footnote{Another argument follows from the fact that $\langle J\rangle\ne0$ on the Coulomb branch (and, of course, similarly on any Higgs or mixed branch). Therefore, we expect $\langle\sO J\rangle\sim\langle\sO\rangle\langle J\rangle\ne0$ in the deep IR. For more details on how to track $J$ to the IR, see \cite{Cordova:2018acb}.}\footnote{In the above argument, unitarity or the existence of a Coulomb branch is crucial. Indeed, consider the free $\CN=2$ Abelian gauge theory with wrong statistics. In this case, $\sO=\varphi$ (since $\varphi^n$ with $n>1$ vanish by Fermi statistics), and $J=\varphi^{\dagger}\varphi$. Therefore, the normal-ordered product clearly vanishes, $\sO J=0$. Next, note that 
\begin{equation}
\sO\epsilon_{ij}\left[Q^i_{\alpha},\left\{Q^j_{\beta},\sO\right\}\right]\sim\varphi F_{\alpha\beta}\sim\epsilon_{ij}\left\{Q^i_{\alpha},\left[Q^j_{\beta},(\varphi^2)\right\}\right]=0~, 
\end{equation}
where we have used the fact that the Lorentz triplet combination of $\epsilon_{ij}\lambda^i_{\alpha}\lambda_{\beta}^j$ vanishes by Bose statistics (recall that the non-unitary gauginos transform as bosons). Therefore, our above argument does not apply to non-unitary theories (including, presumably, the more general ones in \cite{Dijkgraaf:2016lym}).}:

\medskip
\noindent
{\bf Claim 2:} In a unitary 4D $\CN=2$ SCFT with a Coulomb branch, we will have at least one $\bar\CC_{0,r(0,0)}$ multiplet for all $r$ corresponding to $\bar\CE_r$ multiplets with primaries that can take a vev on the Coulomb branch. Moreover, given any two Coulomb branch multiplets, $\bar\CE_r$ and $\bar\CE_{r'}$, we will have at least one $\bar\CC_{0,r_1+r_2-1(1,0)}$ multiplet.

\medskip
\noindent
As a result, we see that the MAD theory has the simplest $\bar\CC_{0,r(j,0)}$ spectrum allowed by unitarity and the existence of a Coulomb branch. If all interacting 4D $\CN=2$ SCFTs have a Coulomb branch, claim 2 can be upgraded to a claim on all interacting 4D $\CN=2$ SCFTs, and the MAD theory would have the simplest $\bar\CC_{0,r(j,0)}$ spectrum of any interacting 4D $\CN=2$ SCFT.

\newsec{A conjecture on the structure of the MAD local operator algebra}
A defining property of the free $\CN=2$ vector multiplet SCFT is that any operator in the theory can be built from products of the various component fields---$\phi$, $\lambda^i_{\alpha}$, $F_{\alpha\beta}$, and conjugates---along with derivatives. Said slightly differently, any local operator in this theory, $\CO$, is in the $(n,m)$-fold OPE of the $\bar\CD_{0(0,0)}\oplus\CD_{0(0,0)}$ multiplets
\begin{equation}\label{FVOPE}
\CO\in \bar\CD_{0(0,0)}\times\cdots\times\bar\CD_{0(0,0)} \times\CD_{0(0,0)} \times\CD_{0(0,0)}:=\bar\CD_{0(0,0)}^{\times n}\times\CD_{0(0,0)}^{\times m}~, \ \ \ \forall\CO\in\CH_{\rm Free Vect.}~,
\end{equation}
where $\CH_{\rm Free Vect.}$ is the Hilbert space of local operators. Note that, to produce the full set of local operators in the theory, we must consider arbitrarily large $n$ and $m$ (this statement follows from $SU(2)_R$ and $U(1)_R$ covariance of the OPE).

Related statements hold in the free hypermultiplet SCFT, where any operator can be built out of the $q^i$, $\psi_{\alpha}$, $\tilde{\bar\psi}_{\dot\alpha}$, conjugates, and derivatives. Similarly to \eqref{FVOPE}, any local operator in this theory, $\CO$, is in the $(n,m)$-fold OPE of the $\hat\CB_{1/2}\oplus\bar{\hat\CB}_{1/2}\simeq \hat\CB_{1/2}\oplus\hat\CB_{1/2}$ multiplets\footnote{Recall that the hermitian conjugate of a $\hat\CB$ multiplet is another $\hat\CB$ multiplet. This latter multiplet may or may not be the same as the original multiplet (in the case of the free hyper, it is not).}
\begin{equation}\label{QQtOPE}
\CO\in \hat\CB_{1/2}\times\hat\CB_{1/2}\times\cdots\times\hat\CB_{1/2}:=\hat\CB_{1/2}^{\times n}\times\hat\CB_{1/2}^{\times m}~, \ \ \ \forall\CO\in\CH_{\rm Free Hyper.}~.
\end{equation}
Again, for the same reasons as in the free vector, we will have to consider $n$ an $m$ to be arbitrarily large in order to obtain all local operators.

Therefore, we see that all local operators at any point on the Coulomb branch of the MAD theory (including points where a single massless hypermultiplet appears) are tightly constrained. It is then natural to ask what happens to the local operator algebra at the origin of the moduli space (i.e., in the MAD theory itself).

Here we have a single Coulomb branch generator, $M\in\bar\CE_{6/5}$, and a single generator of the Schur sector, $J^{11}_{+\dot+}\in\hat\CC_{0(0,0)}$. These operators are clearly in different multiplets. Now, as we have argued, the way to interpret our results on the $\bar\CC_{0,r(j,0)}$ multiplicities in \eqref{Cclaim2} is that these representations are generated in the $\CE_{6n/5}\times\hat\CC_{0(0,0)}$ and $\CE_{6n/5}\times\CE_{6n'/5}$ OPEs. Since there are no other allowed $\bar\CC_{0,r(j,0)}$ multiplets (when, a priori, there could have been infinitely many), we find an infinite amount of evidence for the following conjecture\footnote{In fact, our result in \eqref{Cbarcontri} is also compatible with the conjecture below, since $\bar\CC_{0,r(j,1/2)}$ appears in OPEs involving just $\bar\CE$ and $\hat\CC_{0(0,0)}$ (see \cite{Song:2021dhu,Xie:2021omd} for a discussion of the relevant selection rules).}:

\newpage
\noindent
{\bf Conjecture (MAD operator algebra):} The algebra of local operators of the MAD theory is contained in the $(n,m,p)$-fold OPE of the $\bar\CE_{6/5}\oplus\CE_{-6/5}\oplus\hat\CC_{0(0,0)}$ multiplets\footnote{The general structure of this conjecture involves products of FCS, anti-FCS, and Schur sector operators. In the case at hand, note that $\hat\CC_{0(0,0)}$ appears in the OPE of $\bar\CE\times\CE$.}
\begin{equation}
\CO\in\bar\CE_{6/5}^{\times n}\times \CE_{-6/5}^{\times m}\times\hat\CC_{0(0,0)}^{\times p}~, \ \ \ \forall\CO\in\CH_\text{MAD}~,
\end{equation}
where $\CH_\text{MAD}$ is the Hilbert space of local operators (including those in long multiplets).

\bigskip
\noindent
As in the case of the theories discussed above, we should consider arbitrarily many products if we wish to generate any local operator of the theory. This conjecture has the added benefit of directly generalizing the situations described around \eqref{FVOPE} and \eqref{QQtOPE} on the Coulomb branch of the MAD theory. It would be interesting to understand if we can generalize the statement of this conjecture to apply to other 4D $\CN=2$ SCFTs (e.g., to other AD theories that heuristically look like abelian gauge theories with mutually non-local massless matter).

\newsec{Conclusions}
In this paper, we have taken two steps toward finding the exact spectrum of short multiplets in the MAD theory: we completed the description of the spectrum of chiral operators by showing the MAD theory has no $\bar\CB$ multiplets. We then used this fact to find the exact spectrum of $\bar\CC_{0,r(j,0)}$ multiplets. We also showed that these results have precise counterparts on the MAD Coulomb branch, and we showed that the MAD theory is indeed maximally simple from these perspectives.

The methods employed in this paper can be straightforwardly  generalized to large classes of AD theories that admit $\CN=1$ Lagrangian  descriptions in the UV. It would be interesting to do this and to understand how our results interact with RG flows between these theories.

As a bonus, these results combine to suggest a conjecture on the global structure of the MAD operator algebra. Clearly, it would be interesting to prove this conjecture (or else to find operators constituting a counterexample), generalize it to other classes of 4D $\CN=2$ SCFTs, and to understand its implications for the conformal bootstrap\footnote{See \cite{Cornagliotto:2017snu,Gimenez-Grau:2020jrx,Bissi:2021rei} for bootstrap studies of the $(A_1, A_2)$ SCFT and related theories.} and QFT more generally. We hope to return to these questions soon.

\medskip 
\ack{We thank A.~Bissi, A.~Hanany, H.-C. Kim, A.~Manenti, T.~Nishinaka, E.~Pomoni, and S.~Razamat for discussions. We are especially grateful to T.~Nishinaka for collaboration on related work. M.~B. thanks the organizers at Technion University for a stimulating environment during the workshop, \lq\lq Higher Form Symmetries, Defects, and Boundaries in QFT" and the organizers at Kings College London for a fantastic workshop, \lq\lq Intersections of String Theory and QFT," where some of these results were presented. The work of C.~B. was partially supported by funds from Queen Mary University of London. The work of M.~B. was partially supported by the Royal Society under the grant, \lq\lq New Aspects of Conformal and Topological Field Theories Across Dimensions." The work of M.~B. and H.~J. was partially supported by the Royal Society under the grant, \lq\lq Relations, Transformations, and Emergence in Quantum Field Theory" and the STFC under the grant \lq\lq Amplitudes, Strings and Duality."}

\newpage

\begin{appendices}

\newsec{The FCS in General 4D $\CN=2$ SCFTs}\label{chiralOPgeneral}
As described in the main text, 4D $\CN=2$ SCFTs have Poincar\'e supercharges $Q_{i \,\alpha}$ and $\bar Q^{i}{}_{ \dot\alpha}$ with $i=1,2$. Such a theory  can also be regarded as a 4D $\CN=1$ SCFT whose Poincar\'e supercharges are $Q_{1 \,\alpha}$ and $\bar Q^{1}{}_{ \dot\alpha}$.   The $\CN=1$ $U(1)_r$ charge, $ r_{\CN=1}$,   is given by
 \be
  r_{\CN=1}=\frac23 r +\frac 43 R~,
 \ee
 where $r$ and $R$ are the $U(1)_r$ charge and the $SU(2)_R$ weight respectively.
 
In the conventions described in the main text around \eqref{N1sub} and \eqref{N1perp}, the quantum numbers of all Poincar\'e supercharges are 
  \beqn \label{Qweight}
Q_{1\alpha}\sim Q^2_{\alpha}\in\left(  \frac12,0\right)_{-\frac12,-\frac12}~,\ \ \ \bar Q^1_{\dot\alpha}\sim\bar Q_{2\dot\alpha}\in\left(0,  \frac12\right)_{\frac12,\frac12}~,\nonumber
\\
Q_{2\alpha}\sim Q^1_{\alpha}\in\left(  \frac12,0\right)_{ \frac12,-\frac12}~,\ \   \bar Q^2_{\dot\alpha}\sim\bar Q_{1\dot\alpha}\in\left( 0,  \frac12\right)_{-\frac12, \frac12}~ .
\eeqn
 Note that acting with any Poincar\'e supercharge increases the scaling dimension of an operator by $\frac12$.

 The $\CN=1 $ chiral operators are  those satisfying\footnote{More generally, we can define an $\CN=1$ chiral operator as one annihilated by $\bar Q^{2}{}_{ \dot\alpha}$ or even a linear combination of $\bar Q^{i}{}_{ \dot\alpha} $.  The resulting set of chiral operators are isomorphic due to $SU(2)_R$ symmetry.}
\be\label{chiral}
\left[\bar Q^{1}{}_{ \dot\alpha},\CO\right\}=0~.
\ee
The  $\CN=1$ chiral ring consists of equivalence classes under the relation $\CO\sim \CO +  \left\{\bar Q^{1}{}_{ \dot\alpha},  \CO'\right]$ (for well-defined $\CO'$). For a non-trivial element of this ring, the scaling dimension is fixed in terms of $R$-symmetry quantum numbers\footnote{These operators are automatically conformal primaries. Indeed, chiral conformal descendants are trivial in the chiral ring.}
 \be\label{chiralDim}
 \Delta=\frac32 r_{\CN=1}= 2R+r~.
 \ee

Since we are interested in $\CN=2$ SCFTs, the $\CN=1 $ chiral operators should also sit in representations of the $\CN=2$ superconformal algebra. These operators and their corresponding $\CN=2$ representations are what we have been calling the theory's FCS. Our claim is then that:

\paragraph{Claim:} A non-trivial $\CN=1 $ chiral operator, $\CO$, should sit in an $\CN=2$ multiplet whose $SU(2)_R$ highest-weight superconformal primary, $\CO^\text{SCP}$, is also a non-trivial  $\CN=1 $ chiral operator; in such a multiplet,  $\CO$ can either be the superconformal primary, $\CO^\text{SCP}$, or its $SU(2)_R$ highest-weight $(Q^{1}_{  \alpha})^n $  descendant (where $n=1,2$).
 
\bigskip
   
\noindent \emph{Proof:} Suppose the superconformal primary does not have a non-trivial $\CN=1 $ chiral operator at any $SU(2)_R$ weight. Then all the chiral superconformal descendant operators in this multiplet are $\bar Q^{1}{}_{ \dot\alpha}$-exact. To understand this statement suppose that we have a non-trivial chiral operator as a descendant. Then, such a superconformal descendant must take the form $f(Q,\bar Q)\CO_p$, where $\CO_p$ is some $SU(2)_R$ component of the superconformal primary (or a combination thereof), and $f(Q,\bar Q)$ is a string of Poincar\'e supercharges that must include at least one $\bar Q^1_{\dot\alpha}\in f(Q,\bar Q)$. The reason is that if $\CO_p$ is not a non-trivial chiral operator, then $\Delta(\CO_p)>2R(\CO_p)+r(\CO_p)$, and $\bar Q^1_{\dot\alpha}$ is the only pair of supercharges with $\Delta<2R+r$ (a non-trivial chiral operator should satisfy \eqref{chiralDim}). Now, we can either anti-commute $\bar Q^1_{\dot\alpha}$ to the left of all supercharges in $f(Q,\bar Q)$ and find that $f(Q,\bar Q)\CO_p$ is trivial in the $\CN=1$ chiral ring or there is a $Q_{1\alpha}\in f(Q,\bar Q)$ supercharge to the left of $\bar Q^1_{\dot\alpha}$. Anti-commuting $\bar Q^1_{\dot\alpha}$ past $Q_{1\alpha}$ can be done at the cost of introducing another term involving $\partial_{\alpha\dot\alpha}$ (this follows from the $\CN=1$ SUSY algebra). Note that in order to have $\Delta=2R+r$, we need an additional $\bar Q^1_{\dot\alpha}$ present (since the net contribution of $Q_{1\alpha}$, the first $\bar Q^1_{\dot\alpha}$, and any other supercharges does not decrease $\Delta-2R-r$ relative to the primary; note that this logic excludes the possible appearance of a second $Q_{1\alpha}$). Since $\bar Q^1_{\dot\alpha}$ commutes with $\partial_{\alpha\dot\alpha}$, we have a $\bar Q^1_{\dot\alpha}$-exact term. This is a contradiction.

As a result, the superconformal primary should have a state, $\CO^\text{SCP}$, satisfying \eqref{chiralDim}
\be\label{chirlDimSCP}
\Delta(\CO^\text{SCP}) =2R(\CO^\text{SCP})+r(\CO^\text{SCP})~.
\ee
It is easy to see this state is highest $SU(2)_R$ weight. Indeed, suppose it is not. Then, applying an $SU(2)_R$ raising operator, $R_+$, leads to a state with $\Delta<2R+r$, which is a violation of unitarity.

Let us now examine which descendants $\CO$ can correspond to. First, note that we need only consider states obtained by an action of supercharges on the $SU(2)_R$ highest-weight state, $\CO^{\rm SCP}$. Otherwise, the primary state has $2R+r-\Delta<0$, and we are back to the situation described in the first paragraph of this proof.

Next, let us consider $\CN=1$ chiral descendants obtained by acting with Poincar\'e supercharges on $\CO^{\rm SCP}$. Clearly, $2R+r=\Delta=\frac12$ is only satisfied by $Q^1_\alpha$. Therefore, we can get candidate descendant states $\CO \sim (Q^1_\alpha)^n \CO^\text{SCP} $, where $n=1,2$ (since these states satisfy \eqref{chiralDim}). These states are, by construction, highest $SU(2)_R$ weight.

In fact, there are no other candidate states: if we act with a $\bar Q^1_{\dot\alpha}$ supercharge we are back in the situation described in the first paragraph. On the other hand, acting with $Q_{1\alpha}$ or $\bar Q^2_{\dot\alpha}$ (which all have $\Delta>2R+r$) requires acting with $\bar Q^1_{\dot\alpha}$ in order to have any hope of obtaining an operator satisfying \eqref{chiralDim}. Therefore, we are again back to the situation in the first paragraph.

We therefore conclude that a non-trivial $\CN=1$ chiral operator can appear at most in the following positions
\bea \label{ChiralO}
\CO^{11\cdots1}_{\alpha_1\cdots\alpha_{2j}}\in (j,0)_{R,r} &\xrightarrow{Q^1_\alpha}& \CO_{\alpha_1\cdots\alpha_{2j}\alpha}^{11\cdots11}\oplus\CO_{\alpha_1\cdots\alpha_{2j-1}\alpha}^{11\cdots11}\in (j+\frac12,0)_{R+\frac12,r-\frac12}\oplus(j-\frac12,0)_{R+\frac12,r-\frac12} \cr&\xrightarrow{(Q^1)^2 }&  \CO_{\alpha_1\cdots\alpha_{2j}}^{11\cdots111}\in(j,0)_{R+1,r-1}~,
\eea 
where $\CO^{11\cdots1}_{\alpha_1\cdots\alpha_{2j}} $ is the highest $SU(2)_R$ weight superconformal primary operator in a multiplet of the FCS: $\bar\CE\oplus\bar\CD\oplus\hat\CB\oplus\bar\CB$ (this list can easily be checked via \eqref{chirlDimSCP}). Note that not all of the states in \eqref{ChiralO} are realized in particular multiplets, because some of the descendants in \eqref{ChiralO} may be affected by the shortening conditions. $\Box$
  
\newsec{Case-by-case analysis of candidate FCS generators}\label{trivialB}
In this appendix, let us rule out all the following candidate FCS generators via a case-by-case argument: 
\begin{eqnarray}\label{remaininglst}
&&X~,\ \lambda^2~,\ \phi qq'~,\ \phi qq~,\ qq\phi\lambda_{\alpha}~,\ \phi\lambda_{\alpha}\lambda_{\beta}~,\ qq\lambda_{\alpha}~,\ qq'\lambda_{\alpha}~,\cr && qq\lambda_{\alpha}\lambda_{\beta}~,\ qq'\lambda_{\alpha}\lambda_{\beta}~.
\end{eqnarray}
Recall from the main text that none of these operators can be primaries of a $\bar\CB$ multiplet and cannot sit in $\bar\CE$ multiplets either.
 
\begin{itemize}
\item{$X$: It cannot be a level-two descendant because then it would be in an $\bar\CE_{13/5}$ multiplet. Suppose it is a level-one descendant in a $\bar\CB$ multiplet. Then, in this case, the primary would have $j=1/2$ and $r=11/10<1+1/2$, which would be inconsistent.}
\item{$\lambda^2$: It cannot be a level-two descendant because then the primary is of type $\bar\CE_2$. If it is a level-one descendant, the primary has $R=1/2$, $r=3/2$, and $j=1/2$. However, this would be a $\bar\CD_{1/2(1/2,0)}$ multiplet, which we know are absent \cite{Buican:2021elx}.}
\item{$\phi qq'$: If it is a level-two descendant, we are back to the case of $\bar\CE_{7/5}$. If it is a level-one descendant, the primary has $R=1/2$, $r=9/10$, and $j=1/2$, but $r=9/10<1+1/2$.}
\item{$\phi qq$: If it is a primary, $r=1+j=1$ and we are in a $\bar\CD$ multiplet (but no such multiplet exists). If it is a level-two descendant, we are in an $\bar\CE_{2}$ multiplet. If it is a level-one descendant, then the primary has $R=1/2$, $r=3/2$, and $j=1/2$. This is again a $\bar\CD$ multiplet.}
\item{$qq\phi\lambda_{\alpha}$: If it is a level-one descendant, then the primary has $R=1$, $r=2$, and $j=1$ and would be of type $\bar\CD$ (since $r=1+j$); similarly, $j=0$ is disallowed. If it is a level-two descendant, then the primary has $R=1/2$, $r=5/2$, $j=1/2$. On $SU(2)_R$ grounds, this must take the form $M^n\phi\lambda_{\alpha}$, but this has $r\ne5/2$.}
\item{$\phi\lambda_{\alpha}\lambda_{\beta}$: If it is a level-one descendant, then the primary has $R=1/2$, $r=17/10$, and $j=1/2$ (it cannot have $j=3/2$ because $r<5/2$). By $SU(2)_R$ and spin considerations, this can only correspond to $M^n\phi\lambda_{\alpha}$. However, this has $r\ne17/10$. This cannot be a level-two descendant since the primary has $R=0$ and would be of type $\bar\CE$ (which we have already ruled out).}
\item{$qq\lambda_{\alpha}$: If it is a level-one descendant, then the primary has $R=1$, $r=9/5$, and $j=0$ ($j=1$ is ruled out since $9/5<1+1$). Based on the $R$ weight and spin, such a primary can only be of the form $M^nqq'$ or $M^n(\phi\lambda_{\alpha})^2$. However, these operators have $r\ne9/5$. If $qq\lambda_{\alpha}$ is a level-two descendant, then the primary has $R=1/2$, $r=23/10$, and $j=1/2$. On $SU(2)_R$ grounds, the only possibility is $M^n\phi\lambda_{\alpha}$, but this has $r\ne23/10$.}
\item{$qq'\lambda_{\alpha}$: If it is a level-one descendant, then the primary has $R=1$, $r=6/5$, and $j=0$ ($j=1$ is ruled out since $6/5<1+1$). Based on the $R$ weight and spin, such a primary can only be of the form $M^nqq'$ or $M^n(\phi\lambda_{\alpha})^2$. However, these operators have $r\ne6/5$. If it is a level-two descendant, the primary has $R=1/2$, $r=17/10$ and $j=1/2$. The only possibility is $M^n\phi\lambda_{\alpha}$, but this has $r\ne17/10$.}
\item{$qq\lambda_{\alpha}\lambda_{\beta}$: If it is a level-one descendant, the primary has $R=3/2$, $r=23/10$, and $j=1/2$ (it cannot have $j=3/2$ because $r<5/2$). By $SU(2)_R$ and spin considerations, this can only correspond to $M^n(qq')(\phi\lambda_{\alpha})$. However, $r\ne23/10$. If it is a level-two descendant, the primary has $R=1$, $r=14/5$, and $j=1$. From spin and $SU(2)_R$ considerations, this must be of the form $M^n\phi\lambda_{\alpha}\lambda_{\beta}$, but we have already shown $\phi\lambda_{\alpha}\lambda_{\beta}$ is trivial in the IR FCS.}
\item{$qq'\lambda_{\alpha}\lambda_{\beta}$: If it is a level-one descendant, the primary has $R=3/2$, $r=17/10$, and $j=1/2$ (it cannot have $j=3/2$ because $r<5/2$). By $SU(2)_R$ and spin considerations, this can only correspond to $M^n(qq')(\phi\lambda_{\alpha})$. However, $r\ne17/10$. If it is a level-two descendant, the primary has $R=1$, $r=11/5$, and $j=1$. From spin and $R$ weight considerations, this must be of the form $M^n\phi\lambda_{\alpha}\lambda_{\beta}$, but   we have already shown $\phi\lambda_{\alpha}\lambda_{\beta}$ is trivial in the IR FCS.}
\end{itemize}

\end{appendices} 

\newpage
\bibliography{chetdocbib}

\end{document}